\documentclass[namedreferences]{solarphysics}

\usepackage[hyperref,optionalrh]{spr-sola-addons}
\usepackage{graphicx}        
\usepackage{color}           

\newcommand{\aap}{    {\it Astron. Astrophys.}}

\chardef\us=`\_

\begin{document}

\begin{article}
\begin{opening}

\title{Rieger, Schwabe, Suess-de Vries: The Sunny Beats of Resonance}


\author[addressref={aff1},corref,email={F.Stefani@hzdr.de}]{\inits{F.}\fnm{F.}~\lnm{Stefani}}\sep
\author[addressref={aff1}]{\fnm{G.M.}~\lnm{Horstmann}}\sep
\author[addressref={aff2}]{\inits{M.}\fnm{M.}~\lnm{Klevs}}\sep
\author[addressref={aff1}]{\fnm{G.}~\lnm{Mamatsashvili}}\sep
\author[addressref={aff1}]{\fnm{T.}~\lnm{Weier}}\sep
\address[id=aff1]{Helmholtz-Zentrum Dresden -- Rossendorf, Bautzner Landstr. 400,
D-01328 Dresden, Germany}
\address[id=aff2]{University of Latvia, Institute for Numerical Modelling, 3 Jelgavas street, 
Riga, LV-1004, Latvia}

\runningauthor{F. Stefani {\it et al.}}
\runningtitle{Rieger, Schwabe, Suess-de Vries: The Sunny Beats of Resonance}

\begin{abstract} We propose a self-consistent explanation 
of Rieger-type periodicities, the  Schwabe cycle, and
the Suess-de Vries cycle of the solar dynamo in terms of resonances of various wave 
phenomena with gravitational forces exerted by the orbiting planets.
Starting on the high-frequency side, we show that the two-planet 
spring tides of Venus, Earth and Jupiter are able to excite
magneto-Rossby waves which can be linked with typical Rieger-type periods. 
We argue then that the 11.07-year beat period of those 
magneto-Rossby waves synchronizes an underlying conventional
$\alpha-\Omega$-dynamo, by periodically changing either the field
storage capacity in the tachocline or some portion of the $\alpha$-effect therein.
We also strengthen the argument that the Suess-de Vries cycle appears as an 193-year
beat period between the 22.14-year Hale cycle and a spin-orbit coupling 
effect related with the 19.86-year rosette-like motion of the Sun around the barycenter.

\end{abstract}
\keywords{Solar cycle, Models Helicity, Theory}
\end{opening}
\section{Introduction}

The goal of this paper is to sketch out a comprehensive 
model of the various periodicities of the solar dynamo that
appear on widely different time scales. These include, on the 
high-frequency end,
Rieger-type periods in the range of a few 
hundred days, then - most 
prominently - the 11-year Schwabe cycle, and finally the longer-term 
200-year Suess-de Vries cycle. While modern solar dynamo theory
has at its disposal a broad suit of models to 
plausibly explain 
those different periodicities (more or less) 
separately \citep{Charbonneau2020}, we aim here at a
self-consistent explanation that relies on various gravitational 
influences of the orbiting planets on the Sun and its dynamo. 
Key to this concept is the claim
that the Schwabe cycle represents a clocked, if noisy,  process 
with a mean period of 11.07 years that is
synchronized by the three-planet spring-tide periods of the 
tidally dominant Venus-Earth-Jupiter
system. We hypothesize that the 
necessary deposition of tidal energy is accomplished,
on Rieger-type timescales, as a resonant excitation of magneto-Rossby 
waves by the corresponding 
two-planet spring tides, mainly during cycle maxima with strong 
toroidal field.
Utilizing this deposited energy, the synchronization of the 
Schwabe cycle
relies then on a parametric resonance of a rather 
conventional $\alpha-\Omega$-dynamo 
with the three-planet beat period
of 11.07 years between the underlying two-planet 
spring tides.

On the low-frequency side, we
will pursue the conceptionally similar idea that the 
Suess-de Vries cycle
emerges as a beat period of 193 years, this time 
between the 22.14-year Hale cycle and some
spin-orbit coupling mechanism related to 
the rosette-shaped motion of the Sun 
around the barycenter of the solar system which, in turn, 
is dominated by the 19.86-year periodic
syzygies of Jupiter and Saturn.

Finally, we will speculate about the  appearance
of the longest-time variabilities which usually go 
under the notion of Eddy and Bray-Hallstatt cycles.
While reiterating the intrinsic tendency of the 
Suess-de Vries cycle to undergo chaotic breakdowns into
grand minima, we will also discuss the possibility that some 
Bray-Hallstatt-type periodicity
might appear by virtue of a stochastic resonance phenomenon 
related to the of 2318-year period of the 
Jupiter-Saturn-Uranus-Neptun system which is also visible in the
barycentric motion of the Sun.

Yet, before entering these complicated issues we have to
give at least a preliminary answer to one elementary question...

\section{Problem? What problem?}

The main problem with the problem at hand is that even its
very existence is not generally accepted. Au contraire, 
a couple of recent 
papers by \cite{Nataf2022}, \cite{Weisshaar2023}, and \cite{Biswas2023}
have vehemently refuted any claim for phase stability or clocking 
of the solar cycle,
by referring to various data sets
from the last millennium.

Interestingly, neither of those authors have taken into account the
complementary indication for clocking 
stemming from the early Holocene. Analyzing two 
different sets of algae-related data, \cite{Vos2004} had 
provided remarkable evidence 
for phase stability over thousand years, with a mean period of 11.04 years,
which is - in view of the implied error margins - barely distinguishable from
the 11.07-year period mentioned above. A little blemish of 
these data - the 
appearance of a few intervening 
phase jumps of 5.5 years - was plausibly attributed 
to the specific optimality conditions of 
algae growth.
In \cite{Stefani2020b} 
the signature of the 
arising ``virtual'' phase jumps was also shown 
to be well distinguishable 
from that of any ``real'' phase jumps the
possible occurrence of which
was several times discussed \citep{Link1978,Usoskin2002}
in connection with the solar 
cycle during the last millennium.

As for the latter, things are even more complicated than for the 
early Holocene period. While the series of solar cycles can be
safely reconstructed from telescopic data back to A.D. 1700, say, the cycle 
reconstruction during the Maunder minimum is already quite questionable,
and even more so in the centuries before.

In frame of his ``Spectrum of Time'' project 
\citep{Schove1955,Schove1983,Schove1984} 
the British scientist D. Justin Schove had compiled an enormous body of 
telescopic, naked eye, and {\it aurora borealis} observations into an uninterrupted 
series of solar cycles going back to A.D. 242 (with some interruptions, 
even back to 600 B.C.). Not surprisingly, though, this ambitious project 
has met profound criticism, the gravest counter-argument referring to his
``9 per century'' rule which automatically leads to a phase-stable solar cycle \citep{Usoskin2023,Nataf2022}.
While this argument is formally correct, it does not 
refute the possibility 
that the solar cycle is indeed clocked 
with an 11.07-year 
period, in which case any auxiliary and occasional application 
of the ``9 per century'' rule would do absolutely no harm to an 
otherwise correctly inferred cycle series.

In this hardly resolvable situation, any
reliable data sets that are independent of Schove's cycle 
reconstruction could play a decisive role and would, therefore,
be highly welcome.
An interesting candidate for such a data set, based on $^{14}$C data from 
tree-rings, was recently published by \cite{Brehm2021}
and further analyzed by \cite{Usoskin2021} who generated 
from them a solar cycle series back to A.D. 970.
While the ambiguities of this series were clearly pointed out 
by \cite{Usoskin2021} (who carefully assigned various {\it quality flags} 
to the individual cycle minima and maxima) it was 
uncritically adopted in a recent publication by \cite{Weisshaar2023}. 
Applying statistical methods of \cite{Gough1981} to distinguish 
between clocked and random-walk processes,
these authors claimed high statistical significance for their finding 
of a non-clocked solar dynamo.

Yet, looking more deeply into the underlying cycle series of 
\cite{Usoskin2021}, \cite{Stefani2023a} identified at least one 
highly suspicious cycle around 1845 for which there is no 
evidence from any other cycle reconstruction.
With this superfluous cycle being plausibly rejected, 
a one-to-one match with Schove's (phase-stable) 
cycle series was evidenced back to 
(at least) 1140, except  one second additional cycle amidst the Maunder minimum (1650) 
which,  giving its low quality factor according 
to \cite{Usoskin2021}, should also 
be taken with a grain of salt.

While we do not go so far as to claim perfect evidence for
a clocked solar dynamo,
we strongly refute the claim of \cite{Weisshaar2023} for a high statistical 
significance of a non-clocked process. By contrast, the likely
one-to-one match of the $^{14}$C data with Schove's series would imply
the latter to be (more or less) correct which, in turn, 
would also impugn any 
criticism of Schove's ``9 per century'' rule.
Moreover, 
it remains to be seen how the arguments
of \cite{Nataf2022} and \cite{Weisshaar2023} against 
clocking could be reconciled with the algae data 
of \cite{Vos2004}.

At any rate, in this article 
we consider the phase-stability of the Schwabe 
cycle as a serious working hypothesis, for which it seems worthwhile  
to find a reasonable physical explanation. If, someday in the future, new 
data show up to give unambiguous evidence to the contrary, we 
will be the first to declare this paper, as well as its 
predecessors \citep{Stefani2016,Stefani2018,Stefani2019,Stefani2020a,Stefani2021,Klevs2023}, 
as misleading and futile.

\section{Rieger}

We start with the spectrum of the tidal forces, computed from
the NASA/NAIF ephemerides DE430  \citep{Folkner2014}. Evidently, the most 
prominent peaks in Figure 1 correspond to the two-planet 
spring tides 
of Venus-Jupiter
(118.50 days), Earth-Jupiter (199.44 days), 
and Venus-Earth (291.96 days), see \cite{Scafetta2022}. 
A few more peaks at yet smaller periods are due to the 
spring tides of Mercury with Jupiter (44.9 days), 
Earth (58 days) and Venus (72.3 days). 

\begin{figure}[h]
\includegraphics[width=0.9\textwidth]{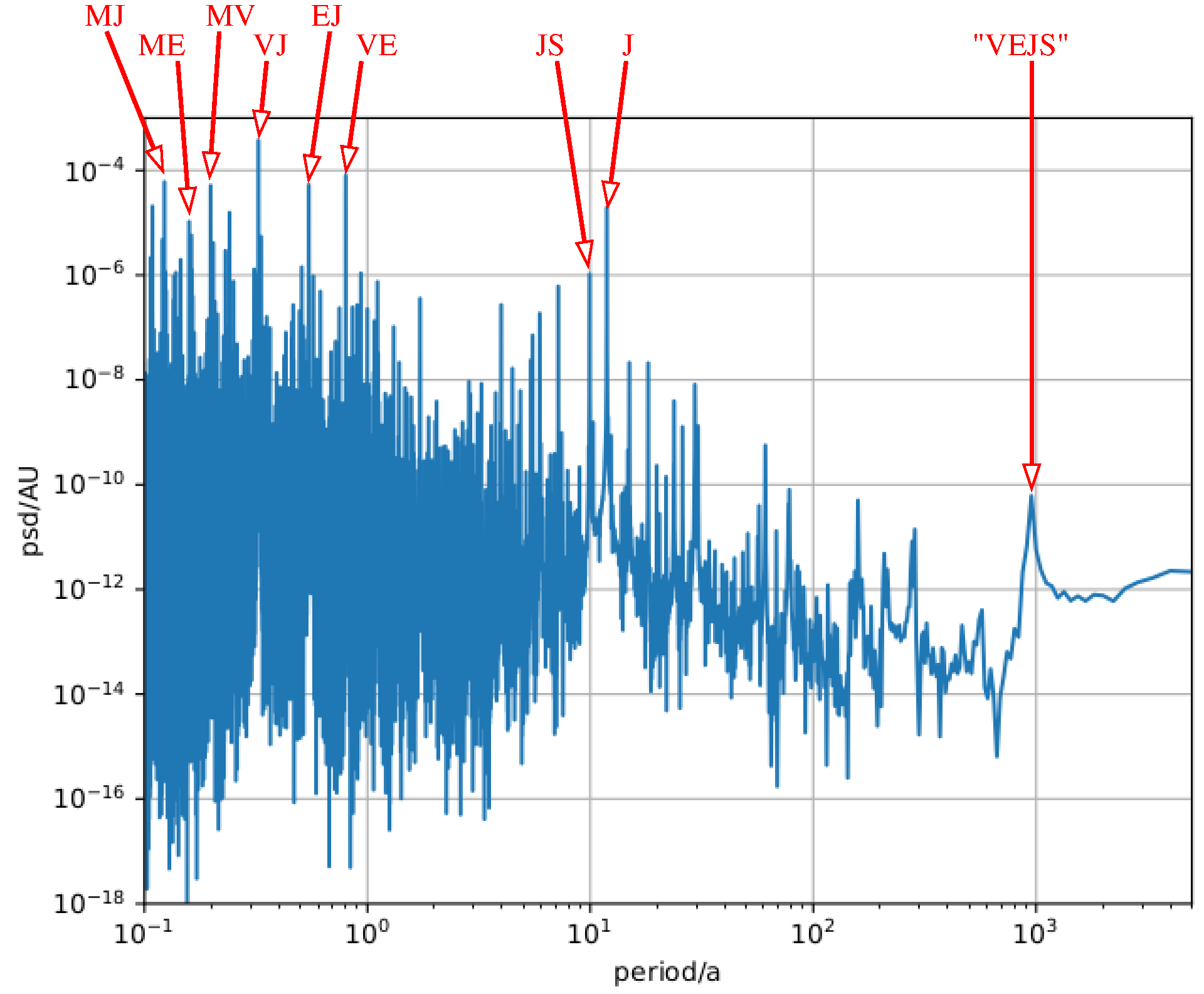}
  \caption{PSD of the tidal forces exerted by all eight planets and 
  Pluto. The main peaks are formed by various two-planet spring-tides except
 that at 11.86 years which stems from Jupiter's orbit, and one at 
 983 years
 which results in a complicated manner from the joint action of
 Venus, Earth, Jupiter and Saturn.}
  \label{Fig:fig1}
\end{figure}

The next prominent peak, at 9.93\,years, corresponds
to the Jupiter-Saturn spring tide which follows from the rosette-shaped, 19.86-year periodic
motion of the Sun around the barycenter of the solar system (which, in turn,
forms the most dominant peak in the PSD of the angular momentum, 
see Figure 3c in
\cite{Stefani2020a}).
Note also the strong
11.86-year peak corresponding to the
orbit of Jupiter.
As noticed by several authors \citep{Nataf2022,Cionco2023},
at this level of {\it linear tidal action} there is no significant 
peak showing up at 11.07 years, a point we will come back to later on.
For an explanation of the last peak at 983 years, 
resulting in a complicated manner from the joint action of Venus, 
Earth, Jupiter and Saturn,
see \cite{Scafetta2022}.

Remarkably, the three dominant peaks at periods corresponding to the 
Venus-Jupiter, Earth-Jupiter and Venus-Earth spring tides belong to the 
band of Rieger-type periods. 
After the initial identification of 
a 154-day periodicity in the occurrence of hard solar flares 
by \cite{Rieger1984}, those periodicities have received 
enormous attention  
\citep{Bai1991,Ballester1999,Dzhalilov2002,
Bilenko2020,Korsos2023}. 

In this context, it had soon be realized that
Rieger-type periods correspond to the 
typical periods of magneto-Rossby waves
during the solar maximum
\citep{Zaqarashvili2010}, a topic which has been 
pursued intensely ever since \citep{Gurgenashvili2016,Marquez2017,Dikpati2018,Gachechiladze2019,Dikpati2020,Dikpati2021a,Dikpati2021b,Gurgenashvili2021}.
Elaborating on this idea
(and also with a side view on a 
similar spring-tide excitation of Rossby-wave 
in the Earth's atmosphere, as discussed recently by \cite{Best2015}),
in this section we use the 
methods of \cite{Horstmann2023} to ask specifically 
what amplitudes 
of magneto-Rossby waves could be generated by the 
individual two-planet spring tides 
of Venus-Jupiter, Earth-Jupiter, and Venus-Earth. 
Further below, we will
consider the nonlinear effects resulting 
from the combination of those individual terms.

Anticipating the typical Rieger-type periods 
involved, we focus here on the most relevant non-equatorial 
magneto-Rossby waves. At this point, a careful distinction must be made between two different types of magneto-Rossby waves. On the one hand, there are slow magneto-Rossby waves, which can reach Gleissberg periodicities \citep{Zaqarashvili2018} and differ fundamentally from classical Rossby waves in that they propagate progradely instead of retrogradely along the latitudes. On the other hand, also retrograde magneto-Rossby waves exist on the classic dispersion branch, which are more similar to hydrodynamic Rossby waves but differ notably in their natural frequencies being right in the range of Rieger periodicities. In the following, we will focus exactly on the latter fast magneto-Rossby waves, which are governed in the framework of the Cartesian $\beta$-plane approximation (with longitudinal coordinate $x$ and latitudinal coordinate $y$) by the following forced wave equation 
\citep{Horstmann2023} for the latitudinal velocity component $v:=v_{\rm y}=v_{\phi}$:
\begin{eqnarray}
	&&\square^2_{v_A}v - C_0^2 \square_{v_A} \Delta v + f_0^2 \frac{\partial^2 v}{\partial t^2} - C_0^2 \beta \frac{\partial}{\partial x}\frac{\partial v}{\partial t} 
	+2 \lambda \frac{\partial}{\partial t}\square_{v_A}v - \lambda C_0^2 \Delta \frac{\partial v}{\partial t} + \lambda^2 \frac{\partial^2 v}{\partial t^2} \nonumber \\ 
	 &&= f_0\frac{\partial}{\partial x}\frac{\partial^2 V}{\partial t^2} - \lambda\frac{\partial}{\partial y}\frac{\partial^2 V}{\partial t^2}  - \frac{\partial}{\partial t}\frac{\partial}{\partial y}\square_{v_A}V \;.
\end{eqnarray}
Here  we use the definitions 
$C_0 = \sqrt{gH_0}$ and $v_A = B_0 / \sqrt{4\pi \rho}$ for the gravity wave and Alfv\'{e}n velocities, respectively (with $g$, $H_0$, $B_0$ and $\rho$ denoting the reduced gravity, height of the tachocline layer, toroidal magnetic field and density of the plasma, respectively),  
and $\square_{v_A} := \partial_t^2 - v_A^2 \partial_x^2$ for the d'Alembert operator with respect to Alfv\'{e}n waves. Further, $f_0 = 2 \Omega_0 \sin(\phi_0)$ and $\beta=(2\Omega_0 /R_0 ) \cos(\phi_0)$ specify the Coriolis and Rossby parameters with $\Omega_0$ and $R_0$ denoting the Sun's angular velocity and the mean radius of the tachocline. Finally, $\lambda$ is an empirical damping parameter and $V$ denotes the external tidal potential. Since the projections of spring-tide envelope potentials onto the $\beta$-plane are quite delicate, we use a simplified potential of a single tide-generating body projected on the $\beta$-plane at mid-latitudes $\phi_0 = 45^{\circ}$
\begin{eqnarray}
	V&=& K\left(\frac{1}{2} +\frac{y}{R_0}\right)\left[1+ \cos\left(\frac{2x}{R_0} + \Omega_{\rm st}t\right)\right] ,
\end{eqnarray}
which oscillates here with the effective spring-tide frequency $\Omega_{\rm st}$ instead of the single planet's orbit frequency. The tidal forcing amplitude is denoted by $K$, for which we use only the value of Jupiter $K = 504\, {\rm cm}\,{\rm s}^{-2}$ for the
sake of comparability with later result. This approach allows us to carry out conservative order-of-magnitude estimates of Rossby wave amplitudes since the tidal peaks associated with the discussed spring-tide frequencies have all similar or even higher powers than Jupiter's peak, see Figure 1. A more accurate computation of
wave excitation by spring tides has to be left for future work.

The response of magneto-Rossby waves to tidal forcing depends
essentially on the damping factor $\lambda$ 
that is mainly (but not exclusionary) 
related to the (turbulent) viscosity 
of the fluid, which in turn is a widely unknown quantity.
In order to get a first clue on the possible reactions, in Figure 2
we consider five different damping factors, 
the higher four ones corresponding to 
$\lambda/\Omega_{\rm st}=10^{-1}...10^{-4}$ and refer to hypothetical lifetimes of freely decaying waves, 
the lowest one (here: $4\cdot 10^{-6}$) 
corresponding to an estimation of boundary layer damping $\lambda \sim \sqrt{\nu \Omega_{\rm st}}/H_0$, which is orders of magnitudes larger than internal damping $\lambda \sim \nu / R_0^2$ in shallow wave systems and often the dominating source of viscous dissipation. 
We have applied the molecular viscosity of $\nu = 10^2$ cm$^2$/s, which is, on the one hand, still conservative in comparison with the even smaller 
estimate 27\,cm$^2$/s of \cite{Gough2007}. On the other hand, in strongly turbulent flows much larger eddy viscosities must be adopted depending on the underlying turbulent scales, which are widely speculative for tachoclinic waves. For that reason this estimation delineates an upper limit for resonant wave excitation.
In this example, we consider 
the 118-day period of the Venus-Jupiter spring tide, 
but use, as said above, 
only the amplitude of Jupiter's tidal forcing.
In doing so, we follow two solar cycles during which the
toroidal field (entering the Alfv\'en speed $v_A$) 
is assumed to have a sinusoidal dependence with an amplitude of 
40 kG. Then, at each magnetic field strength, we search for that 
gravity-wave velocity $C_0$ at which the wave response becomes maximum (Figure 2b). 
This velocity corresponds to a certain height within the tachocline
(with higher velocities corresponding to lower heights and vice versa).
While, quite generally, the amplitude of the magneto-Rossby wave grows with 
the magnetic-field  strength, we observe a strong dependence on the damping parameter
$\lambda$. For the lowest three damping rates, the triggered 
velocity can reach 
values between a moderate 10 cm/s and a whopping
20\,m/s which is indeed remarkable.

For the moment we ignore the principle possibility that the
damping factor might still increase with the 
magnetic field strength, which would play an additional role
in the resonance effects to be discussed later.

\begin{figure}
\includegraphics[width=0.9\textwidth]{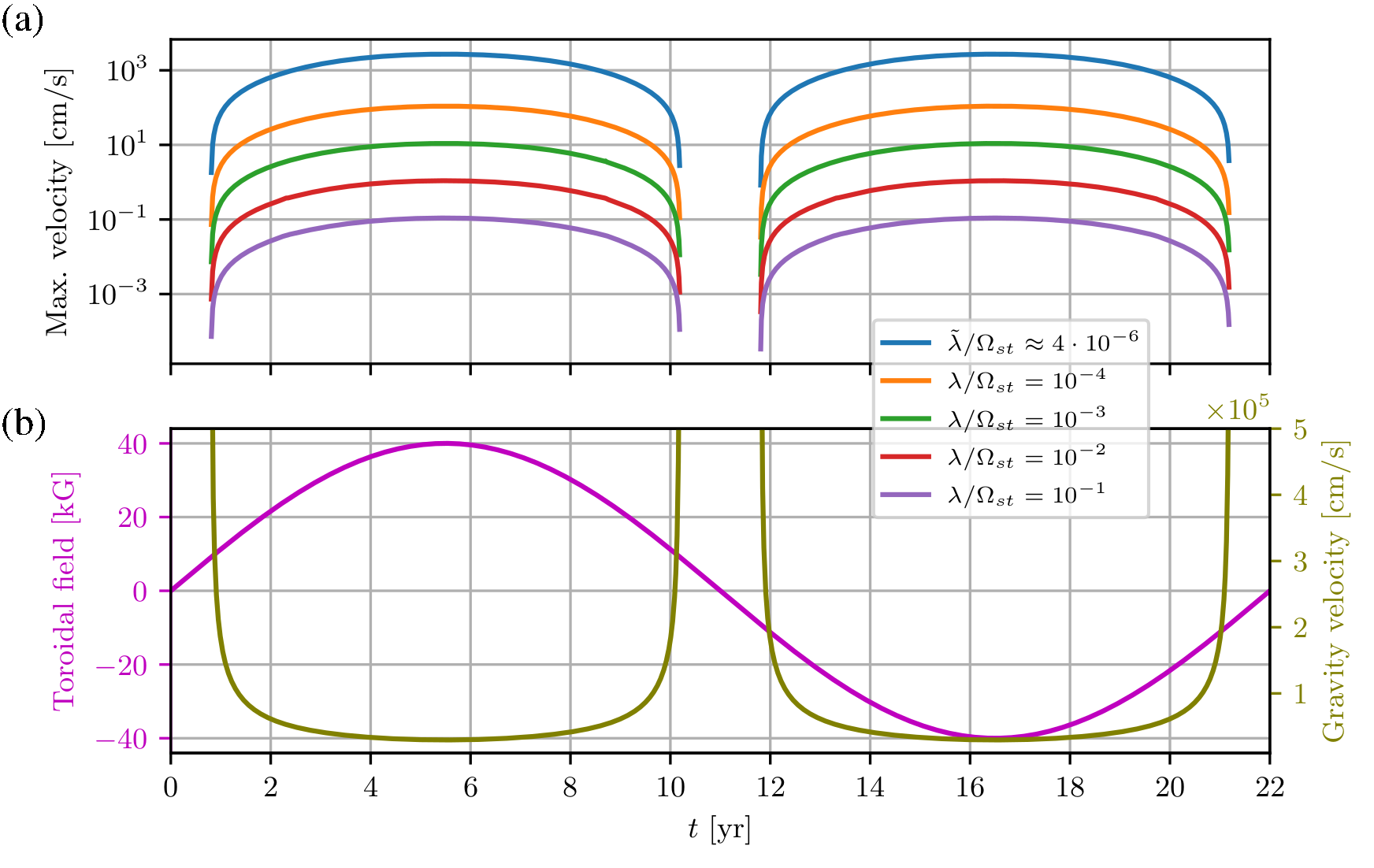}
  \caption{(a) Maximum velocity
  of the magneto-Rossby wave, induced by a tidal-force amplitude of Jupiter, 
  but for an assumed forcing period of 118 days (Venus-Jupiter spring-tide), 
  in dependence
  on an artificial 22-year cycle of the toroidal magnetic field.  
  (b) For each field strength, there is a different gravity-wave 
  velocity, corresponding to 
  a certain depth within the tachocline, which leads to the
  optimum reaction according to (a).
  The response is computed for five different damping parameters, the smallest one corresponding to a viscosity of 100 cm$^2$/s.}
  \label{Fig:fig2}
\end{figure}

Figure 3 shows the respective outcomes for the forcing period 
of 199\,days of the Earth-Jupiter spring-tide.
Evidently the results are not much different from those of Figure 2, 
except that the reached velocities are now a bit 
higher\footnote{Interestingly,
this  enhancement of the wave response with the 
period would be roughly 
compensated by the
slightly lower spring-tides of the Earth-Jupiter 
system}.

\begin{figure}
\includegraphics[width=0.9\textwidth]{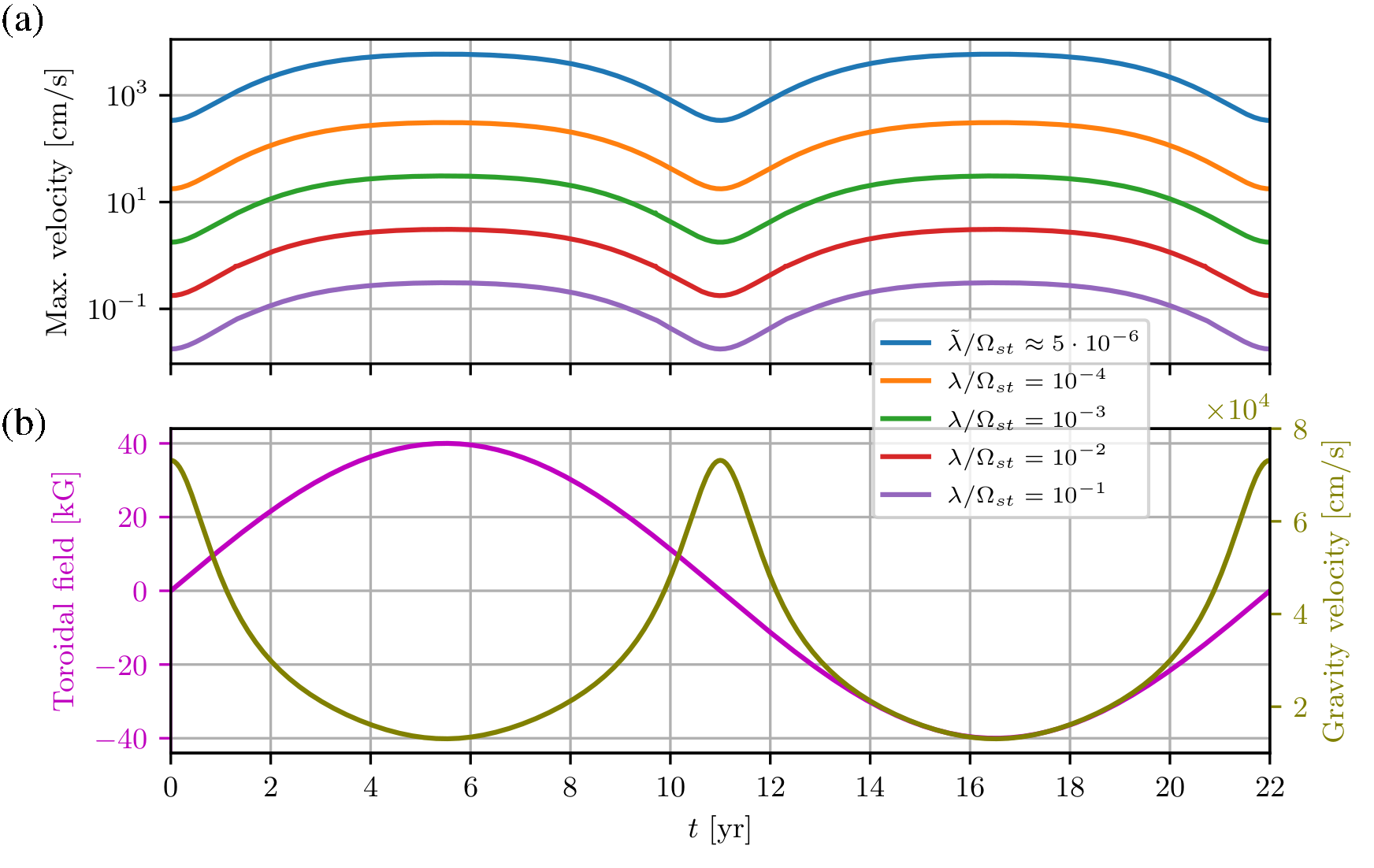}
  \caption{Same as Figure 2, but for an assumed forcing period of 
  199 days (Earth-Jupiter spring tide).}
  \label{Fig:fig3}
\end{figure}

Things are changing, however, when we go over to the 
292-day spring-tide period of Venus-Earth (Figure 4). 
While the excited wave velocities (reaching up to 100\,m/s) 
are generally 
still higher than in Figure 3, at the highest
magnetic fields we observe now
a breakdown of the wave excitation. Evidently, at 
this forcing period we cross the maximum field strength 
beyond which there is no layer of the tachocline 
where the waves can be excited (at least for the parameters used 
in our model). 
Further below, when treating the synchronization problem of the dynamo, 
this behaviour will play a role in justifying the field-dependent 
resonance term for the periodic part of the 
$\alpha$-effect.

\begin{figure}
\includegraphics[width=0.9\textwidth]{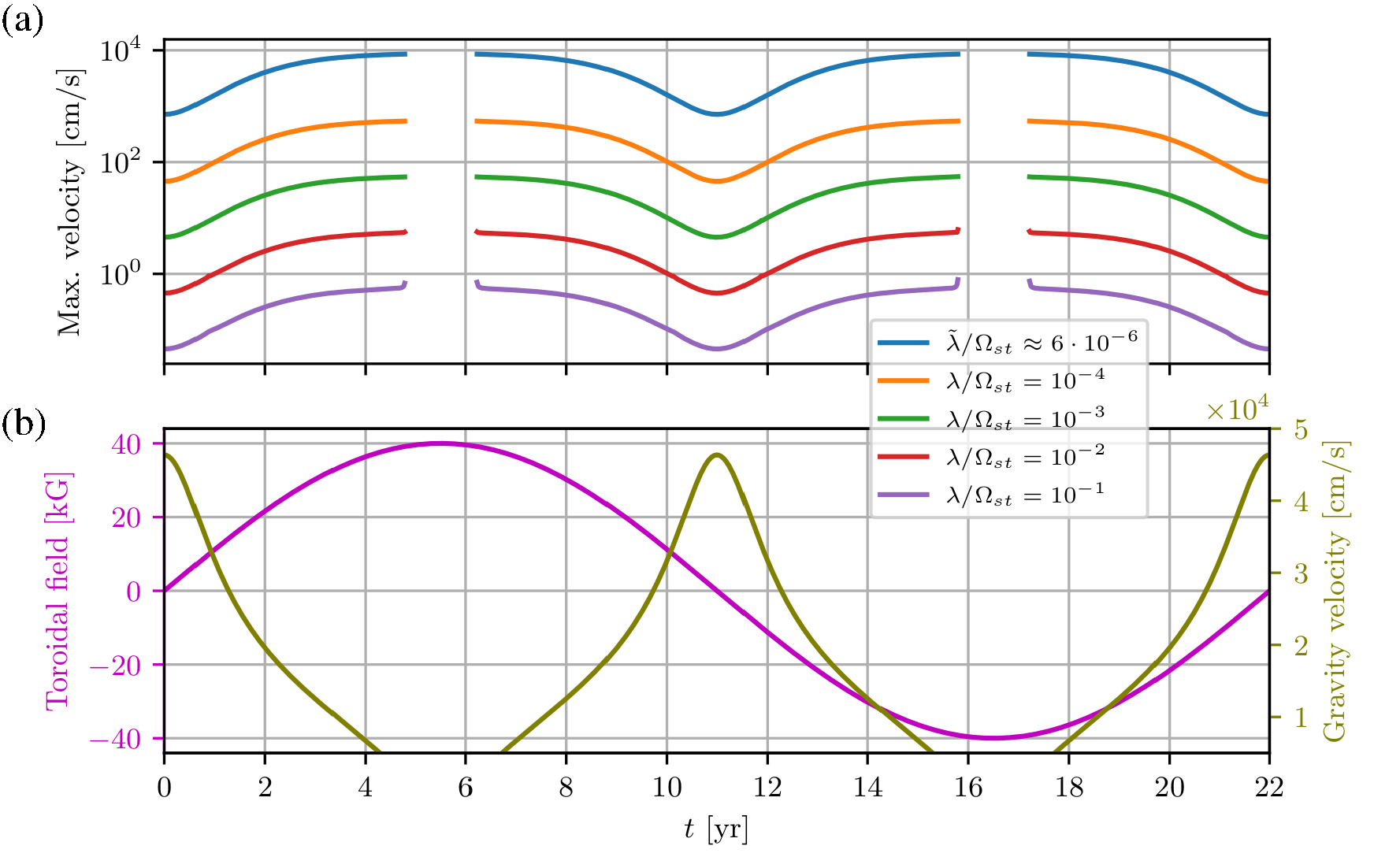}
  \caption{Same as Figure 2, but for an assumed forcing period of 
  292 days (Venus-Earth spring-tide). Note the breakdown 
  of the wave excitation for the {\it strongest} magnetic fields.}
  \label{Fig:fig4}
\end{figure}

At this point, we add another Figure 5 related to the
tidal effects of Mercury. It had been noticed previously
\citep{Hung2007} that the tidal action of Mercury is not much 
weaker than that of Earth, making its omission in 
synchronization models always a bit suspicious. 
As shown now in Figure 5, at the Mercury-Venus spring tide of 72 days
(which leads to relatively modest wave amplitudes anyway)
we get a similar dying-out effect as observed 
for the case of 292 days, but now for the {\it weakest} magnetic fields.
In the extreme case of the Mercury-Jupiter spring-tide period of 45 days,
we have checked that there is no wave excitation at all 
(not shown here).
Obviously, for both too long and too short 
tidal forcings we face the situation that the Sun 
can simply not ``understand'' what the planets are ``telling'' it
since its resonance-ground - in the form of retrograde
magneto-Rossby waves - does not function anymore.

\begin{figure}
\includegraphics[width=0.9\textwidth]{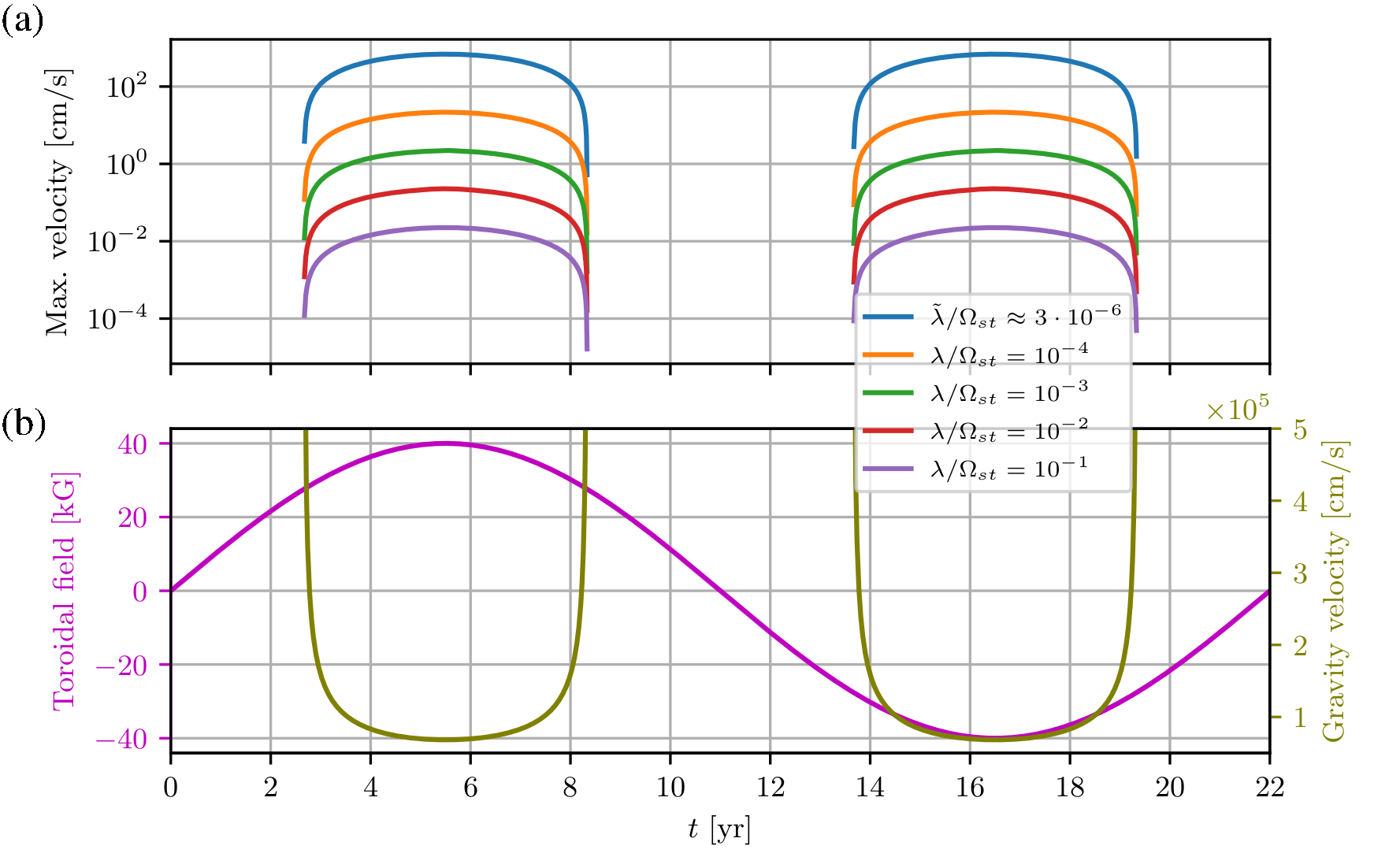}
  \caption{Same as Figure 2, but for an assumed forcing period of 
  72 days (Mercury-Venus spring tide). Note the breakdown 
  of the wave excitation for the {\it weakest} magnetic fields.}
  \label{Fig:fig5}
\end{figure}

What we have learned so far is that the 
three two-planet spring-tide periodicities of the
tidally dominant planets Venus, Earth and Jupiter 
fit amazingly well to the natural periods
of magneto-Rossby waves in the tachocline.
On the very high-frequency side (applying to 
Mercury-Jupiter) waves are
not being excited at all, while on the low-frequency side
(Venus-Earth) the waves were shown to die out 
when the magnetic field becomes too strong.
Moreover, the weakening of the spring-tides 
of Earth-Jupiter (199 days) and Earth-Venus (292 days) 
compared to that of Venus-Jupiter (118 days) is
a sort of compensated by the increasing reaction
of the triggered magneto-Rossby wave at 
those longer periods. 
This effect will become of relevance further below
when we discuss potential non-linear effects of the
three superposed waves.

For the moment, however, it is 
tempting to ask whether the identified spring-tide 
periods do actually show up in the solar data.
While, on the first glance, the very 154-day 
period that was once found by 
\cite{Rieger1984} does not fit 
to any of the tidal periods just discussed, other data look more 
promising. In his spherical harmonic decomposition
of the solar magnetic field, \cite{Knaack2005}
had identified quasi-periodicities grouped 
around 300-320 days, 220-240 days, 170 days, and 
100-130 days. In their analysis of the total solar irradiance (TSI) 
during cycle 23, \cite{Gurgenashvili2021}
had seen two significant 
maxima around 115 days and 180 days. 
Most interestingly, in cycle 24 a significant Rieger
peak close to 150\,days showing up in the VIRGO and SATIRE-S TSI 
data was accompanied by a strong and clear peak at 195 days
in the sunspot area data (their Figure 2).
A similar distinction appeared when comparing the sunspot areas 
in the Northern and Southern hemisphere (their Figure 3). 
With view on such ambiguities, it is not 
completely clear whether the attribution of
185-195-day periods to weak cycles and of 155-165-day 
periods to strong cycles 
\citep{Gurgenashvili2016} always applies.
 
One should also keep in mind that the link
between the intrinsic periods of the (tidally triggered) 
waves and the observable periods might be a bit
{\it indirect}.
One of the intervening factors to consider 
is the rise time of flux tubes from the tachocline to the 
solar surface whose field-dependence is highly nontrivial, 
and, very likely, non-monotonic (see Figure 8 
of \cite{Weber2011}).
With only a very few Rieger-type periods covering the
short strong-field interval of a cycle, 
such a field-dependence of the rise times 
could easily smear out any sharp 
periods of the underlying waves
on the tachocline level
(see, e.g.,  Figure 1 of \cite{Gurgenashvili2016}).
In this respect it is interesting to note that the typical Rieger
period of some 155 days, say, lies quite 
in the middle between the
118 day of the Venus-Jupiter spring tide the 199 days of the
Earth-Jupiter spring-tide, so that this signal might be
fed by a rise-time related modification of both periods.

Apart from the very temporal period, the azimuthal 
dependence of the observed quantities represents 
another clue when it comes to the identification 
of magneto-Rossby waves.
While in our model 
the latter are supposed to have the $m=2$ azimuthal 
dependence as the tidal forcing, the corresponding 
dependence for the observed data is less clear,
despite the fact that active longitudes should be related
to the dominant wavenumbers of the tachocline bulges that
contain toroidal fields \citep{Dikpati2018}.
In many data \citep{Gyenge2016,Gyenge2017,Dikpati2018},
one can find both $m=1$ and $m=2$ contributions, 
typically with some dominance of the $m=1$ part
(see also \cite{Raphaldini2023}).
The latter, however should not be considered as an 
exclusionary argument against a
dominant $m=2$ wave at the tachocline level,
since threshold effects like the launching of flux tubes 
are certainly sensitive to slight symmetry breakings.
Most interesting results on the azimuthal dependence
have been obtained by \cite{Bilenko2020}: distinguishing between
zonal ($m=0$) and sectorial harmonics ($m=n$), this author
had found, in cycle 22, a dominant $m=2$ signal with 
a period of 170-190\,days, and in cycle 23 a corresponding $m=2$ 
structure in the range of 160-200\,days.

While, therefore, neither the temporal nor the azimuthal
dependence of the observations give perfect evidence
for tidally-triggered magneto-Rossby waves, they both 
entail sufficient indications that may 
justify further data analyses to check this hypothesis.

Another feature to be shortly mentioned here is the 
appearance of longer periods such as 
1.3\,years \citep{Howe2000} or 1.7\,years \citep{Korsos2023}
which both are definitely beyond the 
range of excitability of retrograde
magneto-Rossby waves as discussed above.
In this respect it seems worthwhile to consider the possibility of 
intervening non-linearities, e.g.,  tachocline nonlinear
oscillations \citep{Dikpati2018,Dikpati2021a}.

Two types of such nonlinear wave couplings were recently studied in 
detail by \cite{Raphaldini2019,Raphaldini2022}.
In either case they involve quadratic interactions between
the waves which might lead to various frequency and 
wavenumber resonances. 
With respect to the latter, we point to
the analysis of \cite{Bilenko2020} who showed 
the occurrence of dominant $m=4$ modes at the solar maximum,
connected with strong equatorially symmetric $m=0$ modes 
(Bilenko's Figure 4d,e). Obviously, such a  combination 
of $m=4$ and $m=0$ modes 
is exactly what would be expected from 
a quadratic interaction of the tidally triggered $m=2$ 
modes as considered so far.

While we refrain here from a detailed computation of 
the full spatio-temporal structure of those nonlinear terms 
(e.g., in equation systems (25-27) or (28-32) of 
\cite{Raphaldini2019}), we restrict our 
attention to the 
temporal dependence of the
square of the sum of three sine-functions, representing 
different magneto-Rossby waves,
\begin{eqnarray}
    s(t)= \left[ \cos\left( 2\pi  \cdot \frac{t-t_{\rm VJ}}{0.5 \cdot P_{\rm VJ}} \right) +\cos\left( 2\pi  \cdot \frac{t-t_{\rm EJ}}{0.5 \cdot P_{\rm EJ}} \right)
    +\cos\left( 2\pi  \cdot \frac{t-t_{\rm VE}}{0.5 \cdot P_{\rm VE}} \right)
    \right]^2
\end{eqnarray}
with the two-planet synodic periods $P_{\rm VJ}=0.64884$\,years, 
$P_{\rm EJ}=1.09207$\,years,
$P_{\rm VE}=1.59876$\,years, and the 
epochs of the corresponding conjunctions
$t_{\rm VJ}=2002.34$,
$t_{\rm EJ}=2003.09$, and
$t_{\rm VE}=2002.83$
being adopted from \cite{Scafetta2022}.

What we then see in Figure 6a is the appearance 
of a beat period of around 1.6 years.
If we were to skip the Venus-Earth
spring tide in Equation (3)
(e.g., for those times of its disappearance when
the magnetic field is too strong, see 
Figure 4), we would 
obtain Figure 6b showing a beat period 
of  291 days  which, in turn, corresponds 
to the (omitted) Venus-Earth spring-tide\footnote{This correspondence can be understood 
by expressing 
the two-planet spring tide periods $P_{\rm ij}$ 
by the individual periods  $P_{\rm i}$ and $P_{\rm j}$.}.
It remains to be seen whether these 
periods might be related to 
the periods of 0.8-0.9\,years and 1.7-1.8\,years 
as observed by \cite{Korsos2023}.

\begin{figure}
\includegraphics[width=0.8\textwidth]{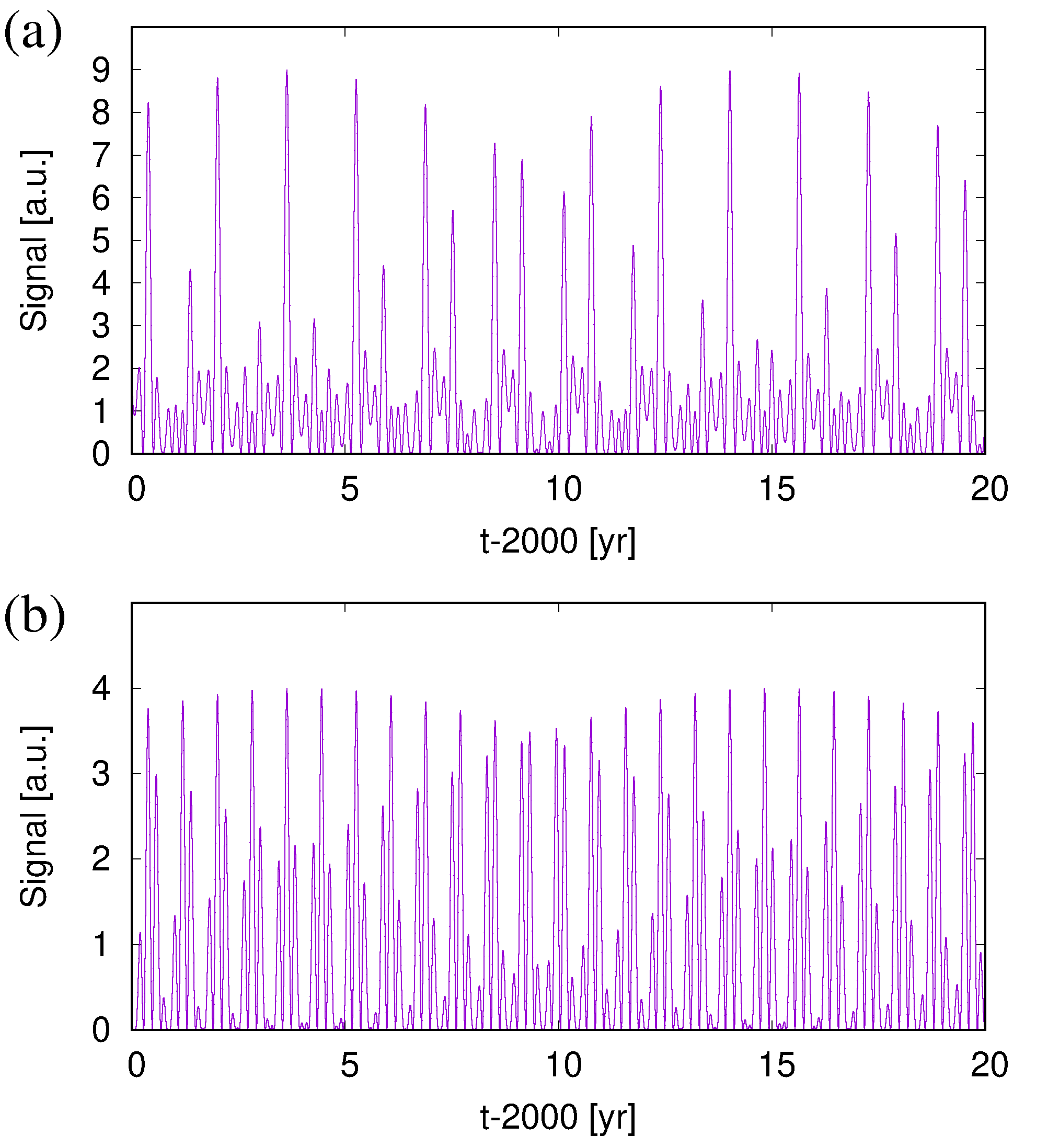}
  \caption{(a) Amplitude of the squared 
  sum of the three individual spring-tide contributions of 
  Venus-Jupiter, 
  Earth-Jupiter, and Venus-Earth. The dominating spiky 
  behaviour corresponds to a period of 
  appr. 1.6 years. (b) The same, but with the
  contribution from the Venus-Earth spring tide being 
  omitted. The dominating 
  spiky behaviour corresponds here to the beat period of 292 days.}
  \label{Fig:fig6}
\end{figure}

\section{Schwabe}
Having seen that the spring tides of two planets are potentially 
capable of triggering magneto-Rossby waves with amplitudes
of the order  of m/s or even more, we will now try 
to figure out how this 
wave energy, once ``harvested'', can be utilized by the solar dynamo 
for synchronizing its Schwabe cycle. The question of whether 
this cycle is synchronized by the planets or not has a long
history, going back to \cite{Wolf1859} and \cite{delarue1872}, 
and being further pursued by \cite{Bollinger1952}, \cite{Takahashi1968}, 
\cite{Wood1972}, \cite{Opik1972}, \cite{Okal1975}, \cite{CondonSchmidt1975},
\cite{Dicke1978},  \cite{Hoyng1996}, \cite{Zaqa1997}, \cite{Palus2000}, 
\cite{DeJager2005}, and \cite{Callebaut2012}. However, 
the amazingly precise correspondence with the three-planet spring 
tides of 11.07\,years was discussed only recently by \cite{Hung2007}, 
\cite{Scafetta2012}, \cite{Wilson2013} and \cite{Okhlopkov2016}. 

Apart from some more exotic 
models relying on changed rates of nuclear fusion in the 
Sun's core \citep{Wolff2010,Scafetta2012}, there are basically 
four
mechanisms which seem suitable to provide a 
discernible effect on the dynamo. 
The first one  assumes that the ellipticity 
of the Sun's barycentric motion leads to
periodic changes of its differential rotation 
\citep{Zaqa1997}.
The
second one relies on the high sensitivity of the field
storage capacity within the weakly sub-adiabatic 
tachocline,  which might easily react even to weak (pressure) 
disturbances by releasing more or less magnetic 
flux tubes
for its rise to the Sun's surface. This concept,
originally  introduced 
by \cite{Abreu2012} (based on \cite{Ferrizmas1994},
was recently tested in a
2D Babcock-Leighton variant of a synchronized 
dynamo model by \cite{Charbonneau2022} 
(relying on the 
time-delay concept of \cite{Wilmotsmith2006}).
Exhibiting
the same type of parametric resonance
as previously observed by \cite{Stefani2019},
this promising result superseded 
the skeptical 
judgment of the synchronization suitability
of Babcock-Leighton type models
that was based on an evidently too 
simple 1D model of \cite{Stefani2018}.

A third potential mechanism for dynamo 
synchronization had
emerged with the observation that
the kink-type, current-driven Tayler 
instability (TI) \citep{Tayler1973,Seilmayer2012} is
prone to intrinsic helicity oscillations 
\citep{Weber2013,Weber2015},
at least for small magnetic Prandtl numbers
(which applies to the tachocline).
While first identified in the 
simplified setting of TI in a non-rotating, full cylinder, 
a similar oscillatory behaviour was recently 
shown to also appear
in a much more realistic 3D-model of the tachocline
(Figure 16 in \cite{Monteiro2023}).
A key observation of \cite{Stefani2016} was then 
that those
helicity oscillations  of the TI (with its 
azimuthal wave number $m=1$) 
are sensitive to entrainment by tide-like ($m=2$) 
perturbations. For the not dissimilar case of 
an $m=1$ {\it Large Scale Circulation} in a 
Rayleigh-B\'enard convection problem this
concept of helicity synchronization by tide-like forcing 
was recently exemplified in a liquid metal 
experiment \citep{Juestel2020,Juestel2022}.  

Slightly modifying the concept of directly forced 
helicity synchronization, 
we focus here on the fourth
possibility that the (tidally triggered) 
magneto-Rossby 
waves themselves give rise to an $\alpha$-effect, 
a possibility that 
was first studied by \cite{Avalos2009}. While 
the $\alpha$-effect for 
stationary Rossby waves is typically zero 
(due to the
90$^{\circ}$ shift of vertical velocity and 
vorticity) these 
authors found a non-vanishing $\alpha$-effects for the case
of drifting waves. This is indeed the scenario
we are dealing with in this paper, and it remains
to be seen how much of the $\alpha$-effect 
{\it is entailed}
in the superposition of the three magneto-Rossby waves
as excited by the two-planet spring tides discussed 
above.

Turning this question upside-down, in \cite{Klevs2023} 
it was asked how much of the synchronized  $\alpha$-effect 
would actually {\it be needed} to entrain the 
entire solar dynamo by parametric resonance. Based on a rather 
conventional 2D $\alpha-\Omega$ dynamo model, including meridional 
circulation, this value was shown to lay in the range of some
dm/s or even less (with the  
resistivity jump between tachocline and convection 
zone being the most decisive, 
yet widely unknown factor).

This consistency between 
the scale of m/s which {\it results} for the velocity of 
tidally excited magneto-Rossby waves
and the scale of dm/s that would be {\it required} for a 
dynamo-synchronizing $\alpha$-effect is indeed
encouraging. Admittedly, it is still a long way to 
corroborate this link in more detail.
For a first glance on this issue, we will again 
restrict ourselves to some simple considerations, leaving a 
detailed computation of the helicity, and the $\alpha$-effect 
connected with it \citep{Avalos2009}, to future work.

The central question in this respect is how the 11.07-year period, 
which is known to have {\it no} significant peak in 
the total tidal potential (see Figure 1 and 
\cite{Okal1975}, \cite{Nataf2022}, \cite{Cionco2023}) might still emerge 
from the interactions 
of the much shorter two-planet spring-tides. 
The classical answer to that question is given in 
Scafetta's formula for the period $P_{\rm VEJ}$ 
of the three-body spring tide \citep{Scafetta2022}
\begin{eqnarray}
    P_{\rm VEJ}=\frac{1}{2}\left[\frac{3}{P_{\rm V}}-\frac{5}{P_{\rm E}} + \frac{2}{P_{\rm J}}  \right]^{-1}
\end{eqnarray}
where $P_{\rm V} = 224.701$\,days, $P_{\rm E} = 365.256$\,days,
and $P_{\rm J} = 4332.589$\,days are the sidereal 
orbital periods of Venus, Earth and Jupiter,
respectively.
The vector $(3,-5,2)$, appearing in Equation (4), can also be written 
as  $(3,-5,2)=3(1,-1,0)-2(0,1-1)$, 
indicating the frequency of the beat being created by 
the third harmonic of the synodic cycle between Venus and 
Earth and the second harmonic of the synodic cycle between Earth and 
Jupiter \citep{Scafetta2022}.

\begin{figure}
\includegraphics[width=0.80\textwidth]{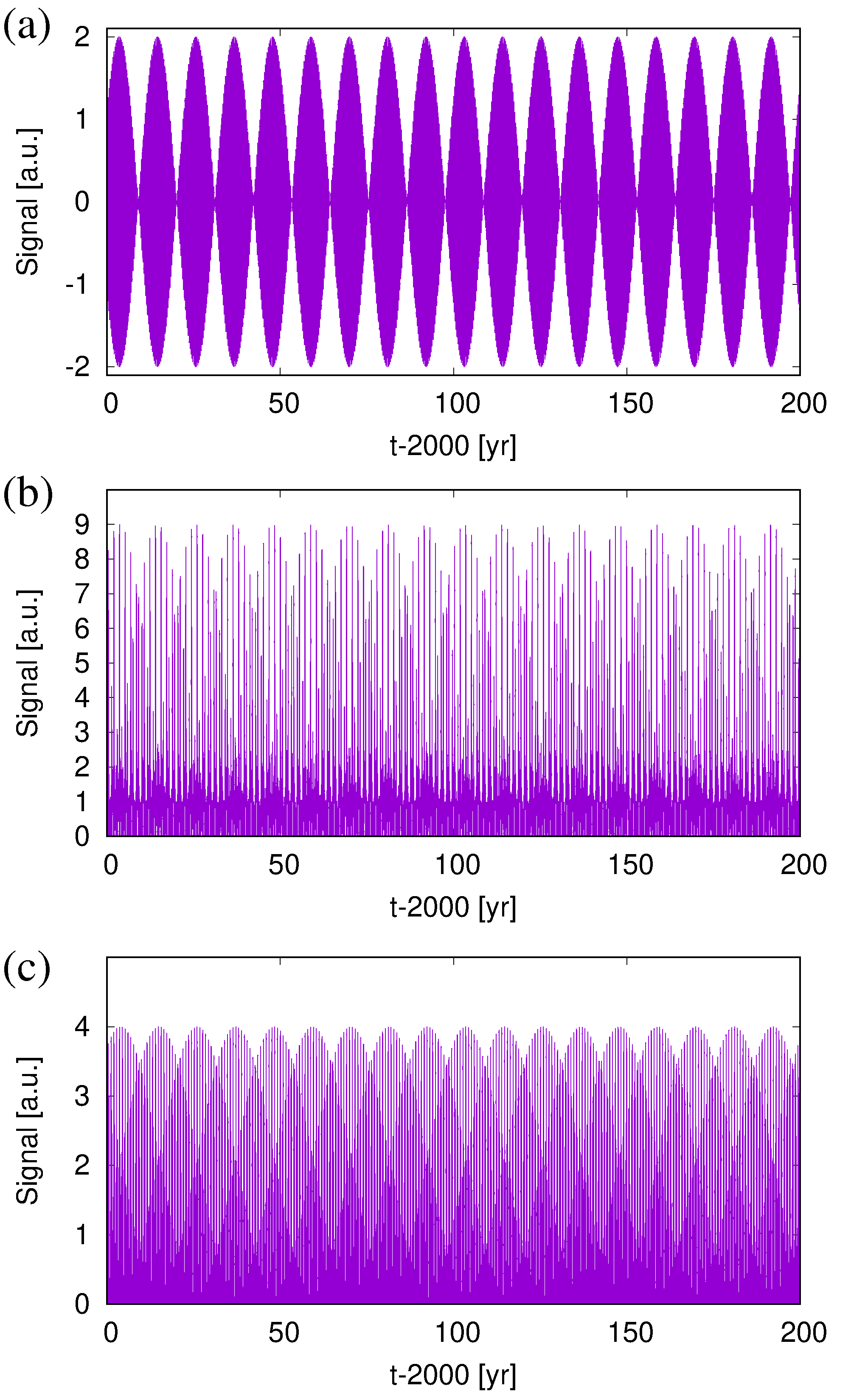}
  \caption{Signal of various combinations of 
  spring tides. (a) Scafetta's equation (5) showing the
  sum of the third harmonic of the synodic cycle between Venus and Earth 
  and the second harmonic of the synodic cycle between Earth and Jupiter. (b) Squared 
  sum of the individual (and equally weighted) 
  spring-tide contributions of Venus-Jupiter,
  Earth-Jupiter, and Venus-Earth according to Equation (3). (c) Same, but
  without  the Venus-Earth contribution (i.e., last term in Equation (3)).
  In either case the three-planet spring-tide period of 11.07 years becomes 
  visible.}
  \label{Fig:fig7}
\end{figure}

The emergence of this beat 
is illustrated in Figure 7a in which we replot the same function
\begin{eqnarray}
    s(t)=  \cos\left( 2\pi  \cdot 2 \cdot \frac{t-t_{\rm EJ}}{0.5 \cdot P_{\rm EJ}} \right)
    +\cos\left( 2\pi \cdot 3 \cdot \frac{t-t_{\rm VE}}{0.5 \cdot P_{\rm VE}}
    \right)
\end{eqnarray}
as in Figure 2A of \cite{Scafetta2022}, showing a clear-cut beat 
period of 11.07\,years. 

Convincing as this curve might look, it is not 
entirely clear how exactly this very particular 
combinations of third and second harmonics 
of the two-planet spring tides 
might be of any physical relevance. An important 
aspect was attributed by Scafetta to the 
feature of ``orbital invariance'', meaning that
the identity $3-5+2=0$ involves that all parts of the 
differentially rotating body feel the same forcing.

Interestingly, the very same 11.07-year beat period
shows up when we only consider the 
square of the sum of the three cosine functions
of Equation (3)
which is indeed expected to occur, e.g., in typical
quadratic functionals of the velocity perturbations, such as, e.g.,  the Eliassen-Palm flux that is well-known in meteorology
\citep{Trenberth1986}, pressure perturbations,
or the helicity.
This is clearly seen in Figure 7b which illustrates again the signal from 
Equation (3), this time, however, for the enlarged interval of 
200 years\footnote{It would be an interesting, if tedious,
exercise in applying the addition theorems for 
trigonometric functions to quantify how much of the 
``ideal'' expression (5) is contained in the
square of three functions according to Equation (3).}. 
And even in case that the Venus-Earth 291-days spring tide 
is neglected (e.g., due to a too strong magnetic field)
it appears that we could obtain the same 11.07 years beat, 
as evidenced in Figure 7c.

However, the latter conclusion turns out to be a bit pre-mature.
With view on synchronized dynamo models such as in \cite{Klevs2023},
we should presumably focus on the axi-symmetric component of the
quadratic term in Equation (3), very likely with some
phase shift of one of the two factors (which would be needed when 
computing the helicity). For the simplest case without phase
shift between the two factors, we show in 
Figure 8a the azimuthal average

\begin{eqnarray}
    S(t)&=& \frac{1}{2 \pi} \int_0^{2 \pi} d\varphi
    \left[ \cos\left( 2\pi  \cdot \frac{t-t_{\rm VJ}}{0.5 \cdot P_{\rm VJ}} +2 \varphi \right) +\cos\left( 2\pi  \cdot \frac{t-t_{\rm EJ}}{0.5 \cdot P_{\rm EJ}} +2 \varphi\right) \right.  \nonumber \\
    &&\,\,\,\,\,\,\,\,\,\,\,\,\,\,\,\,\,\,\,\,\,\,\,\,\,\,\,\,\,\,\,\,\,\,\,\,
    +\left.  \cos\left( 2\pi  \cdot \frac{t-t_{\rm VE}}{0.5 \cdot P_{\rm VE}} +2 \varphi \right) \right]^2
\end{eqnarray}
wherein the phase $2\varphi$ reflects the $m=2$-character of the tidally
triggered waves. As we can see, the 11-07-yr period is still 
visible in this integral, just as it was in Figure 7b. 
Yet, things are changing when we skip in Equation (6) 
the last term in the sum, corresponding to the Venus-Earth spring tide.
In this case, the 11.07-year period disappears completely, so that
Figure 8b looks very different from Figure 7c. In the Appendix,
we will explain these different behaviours 
in terms of the azimuthal dependencies 
of the square under the integral in Equation (6), and relate them to 
Scafetta's notion of ``orbital invariance'' which refers to 
an action that is ``being simultaneously and coherently seen by any 
region of a differentially rotating system'' \citep{Scafetta2022}.

\begin{figure}
\includegraphics[width=0.80\textwidth]{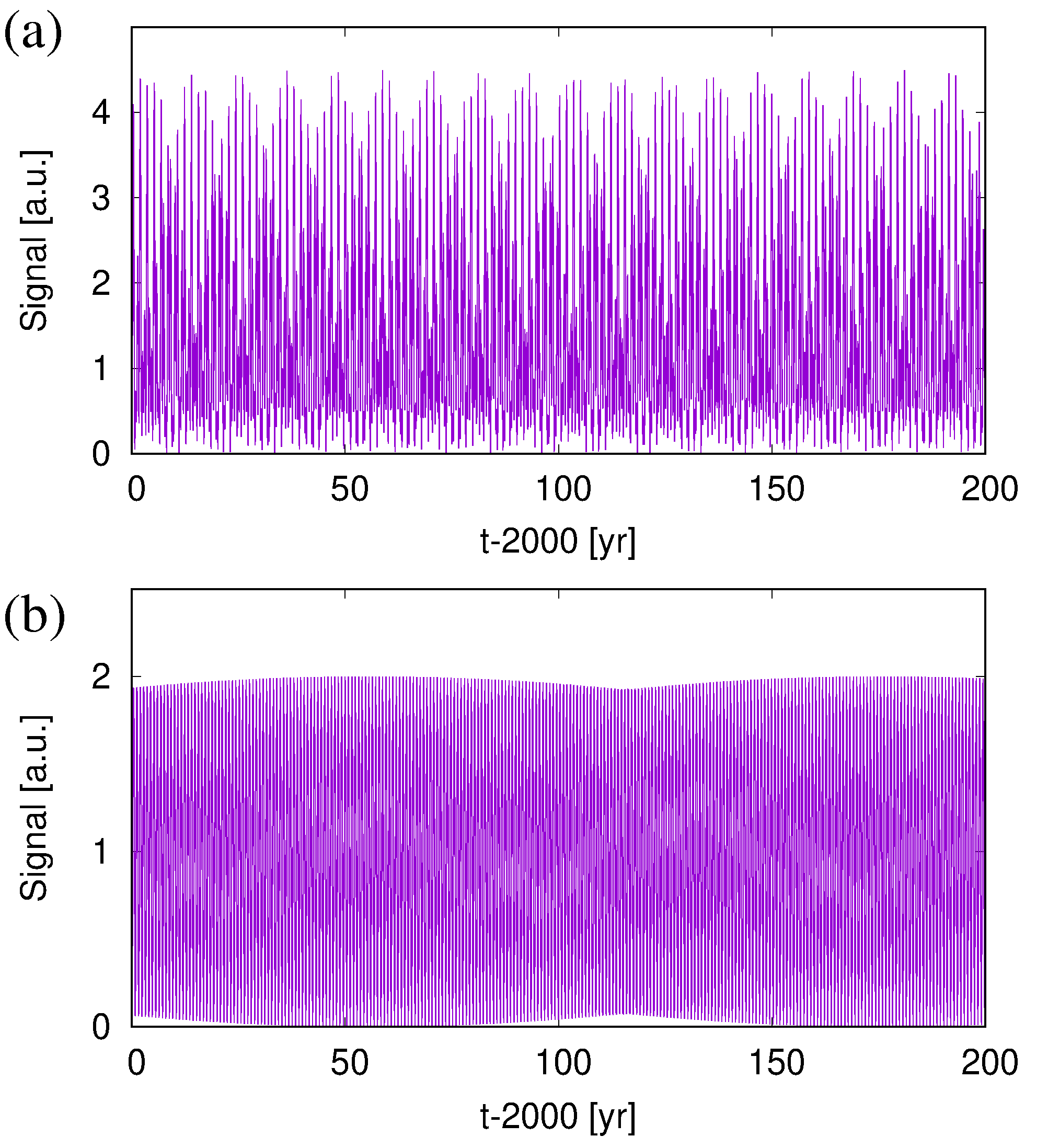}
  \caption{ (a) Azimuthally averaged squared 
  sum of the individual
  spring-tide contributions of Venus-Jupiter, and
  Earth-Jupiter and Venus-Earth according to Equation (6). (b) Same, 
  without  the Venus-Earth contribution (i.e., last term in the sum 
  of Equation (6)).
  The three-planet spring-tide period of 11.07 years remains visible
  only in case (a)}
  \label{Fig:fig8}
\end{figure}

We refrain here from corroborating more details on how the 
superposition of three (or two)  
tidally triggered magneto-Rossby waves 
would generate helicity, or an $\alpha$-effect, for that matter.
In the very simplest case, the azimuthally averaged helicity 
would integrate to zero, when the second factor 
under the integral in Equation (6) is endowed with an
additional phase shift of 90$^{\circ}$ that comes into play when
taking the curl of the velocity
(though a bit trivial, this vanishing is also illustrated in 
the Appendix).
However, as shown by \cite{Avalos2009}, for more complicated
geometries some value of $\alpha$ still remains, and there are good 
reasons to expect it to have a similar 11.07\,year modulation as that 
in Figure 8a.

At any rate: with view on the parametric resonance of the 
underlying conventional 
dynamo having a natural period not too far from 22\,years, 
the dynamo will
react preferably on perturbations with periods around 
11\,years (see Figure 10 in \cite{Stefani2019}, Figure 
10 of \cite{Charbonneau2022}, and Figure 6 in \cite{Klevs2023}).
For this resonance, azimuthally averaged quantities such as in
Figure 8a will presumably play the dominant role.

A this point, the discerning reader will have 
noticed a slight, but important conceptual
shift.
In all previous works 
\citep{Stefani2016,Stefani2018,Stefani2019,Stefani2021,Klevs2023}, 
the 11.07-year periodicity was implemented 
as a resonance term with the generic 
field-dependence $\sim B^2/(1+B^4)$ that
was interpreted in terms of an optimal reaction, 
{\it on the Schwabe time-scale}, of 
the TI to the amplitude of the field. 
Now we re-interpret the very same resonance term as 
an optimal response of the magneto-Rossby waves 
to the magnetic field
on the {\it Rieger-type time-scale}. As was shown above in
Figure 4,  not only too low fields (as in 
Figure 5) 
prevent the wave from being excited, but 
too high fields do so as well.
This is all the more plausible as magnetic fields might also
contribute to the damping factor $\lambda$, 
resulting in another reduction of the wave amplitudes 
with increasing field.

This new interpretation opens up the possibility that 
very weak magnetic fields could possibly 
prevent the excitation of Rieger-type oscillations,
which might 
even lead to a loss of synchronization. In this 
respect it would be interesting 
if  the hypothesized  additional cycle around 1650 
\citep{Usoskin2021}, i.e. deep in the
Maunder minimum, could be confirmed by independent 
data.  A closely related question refers to the
observation of \cite{Reinhold2020} that the 
Sun's activity appears to show a significantly 
lower variability than 
other stars with near-solar temperatures and rotation periods.
In this regards, and given some interesting 
applications in other 
fields \citep{Pontin2014},
it is not unlikely that 
this lower variability, or noise level, 
indeed results from parametric 
resonance.

\section{Suess-de Vries}

There is some tradition of linking long-term temporal variabilities
of the solar cycle to correspondingly long periods of 
planetary influences \citep{Jose1965,Charvatova1997,Landscheidt1999}. 
One of the
most detailed attempts in this direction was that of
\cite{Abreu2012} who had identified quite a couple
of common periodicities. Yet, their approach was soon 
criticized, for various reasons, by \cite{Cameron2013} and \cite{Poluianov2014}.
More recently still, \cite{Charbonneau2022} put into question 
the entire approach 
of linking long-term cycles with correspondingly long
periodicities of the planetary system by arguing 
that ``(b)ecause the low frequency response is internal to the 
dynamo itself, it is similar for incoherent stochastic forcing and 
coherent external periodic forcing''. This lead him to the
conclusions ``that if external forcing of the solar dynamo
of any origin is responsible for the centennial and millennial 
modulations of the magnetic activity cycle, the required forcing 
amplitudes should be 
well beyond the homeopathic regime in order to produce a detectable 
signature in the presence of other sources of modulation, whether deterministic or stochastic.''

\begin{figure}
\includegraphics[width=0.8\textwidth]{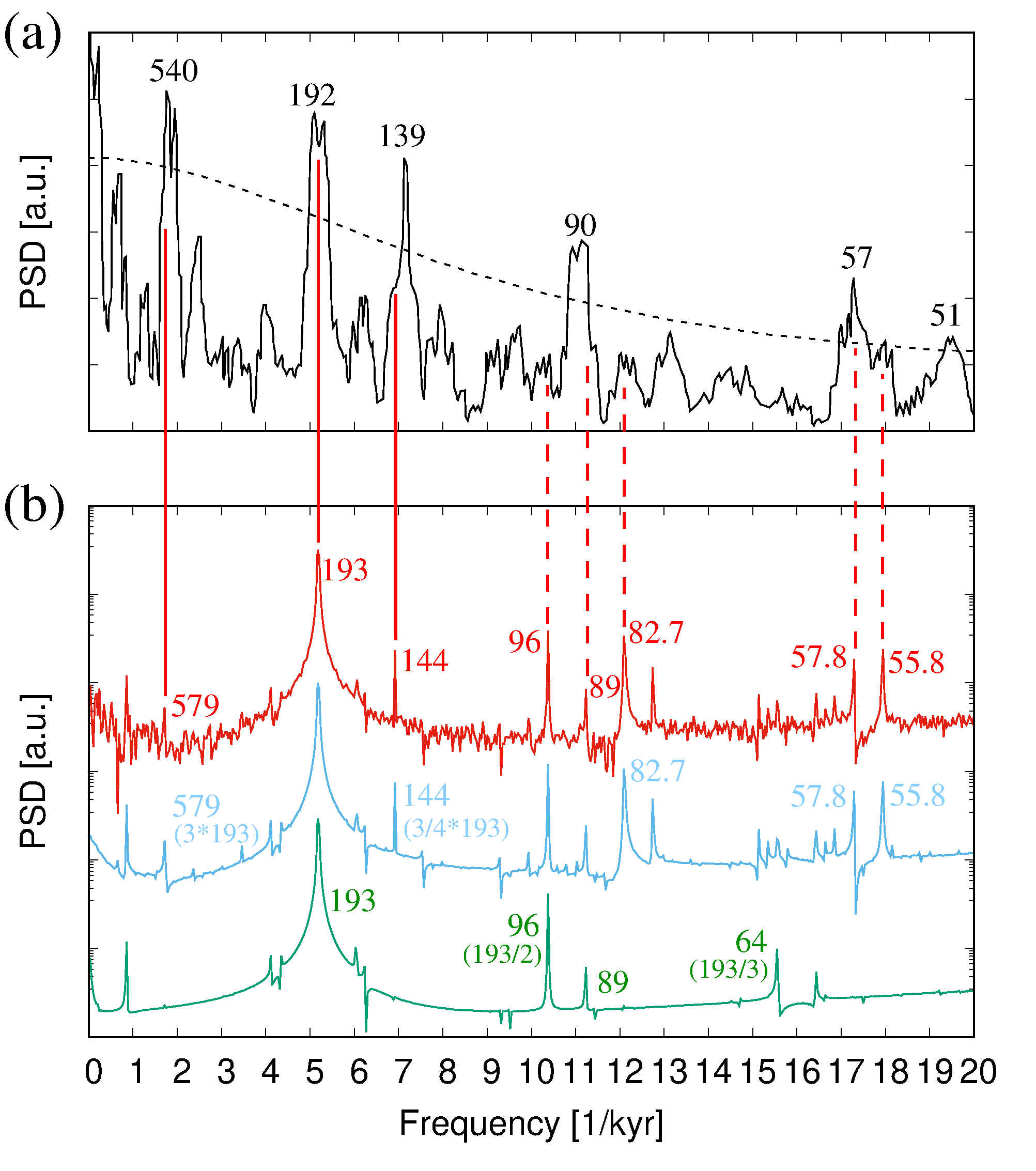}
  \caption{Comparison between the spectra of 
  climate-related observations and
  numerical simulations. (a) Spectrum of
  varved lacustrine-sediment data from Lake Lisan,
  adapted from Figure 3A of \cite{Prasad2004}. The dashed black
  line indicates the 95\,\% confidence interval, the significant
  periodicities are labeled in years. The ordinate axis [a.u.]
  is linearly scaled.
  (b) Spectra of a 1D $\alpha-\Omega$-dynamo model
  (here, the ordinate axis is logarithmic). The green curve (only 
  Jupiter and Saturn included in the computation of the
  orbital angular momentum) and 
  the blue curve (all planets included) correspond to 
  those in Figure 9 of \cite{Stefani2021} (with periods just 
  replaced by frequencies). The red curve is similar to the blue 
  one, but incorporates some noise in the 
  $\alpha$-term. For more details, in particular regarding 
  the origin 
  of the indicated periodicities, see the main 
  text. Note the good agreement of the main peaks for the observations 
  (a) with those of the numerical predictions (b), in particular 
  regarding the sharp Suess-de Vries cycle at 193\,years.}
  \label{Fig:fig9}
\end{figure}

But what if the external forcing acts on much shorter
than centennial and millennial periodicities, and the latter
show up only as beat periods of the former?
This concept, which  traces back to ideas of 
\cite{Solheim2013} and \cite{Wilson2013}, was 
pursued in \cite{Stefani2020a} and \cite{Stefani2021},
where a combination of a primary 11.07-year forcing of the 
$\alpha$-effect with the 19.86-year period 
related to the barycentric motion of the Sun led to the emergence
of an 193-year period. In view of very similar periods 
identified  by \cite{Prasad2004}, \cite{Richards2009}, \cite{Luedecke2015}, and 
\cite{Ma2020} this might indeed be 
the Suess-de Vries cycle, although 
its period is often quoted  with the  slightly larger 
value of 200 or even 208\,years. More complicated still, both in 
terms of observation and modelling, is the case
of the Gleissberg cycle(s) whose occurrence was, in 
Figure 9 of \cite{Stefani2021}, attributed partly to the second and
third harmonics of the 193-year period, and partly to the additional
effects of Saturn-Uranus and Saturn-Neptune. 

In Figure 9, we compare these results with the spectrum of
climate-related varved-sediment data from Lake Lisan, Dead Sea Rift, 
as obtained by \cite{Prasad2004} for the time interval
26.2-17.7 (calendar) ka (Figure 9a). The green and blue curves 
in Figure 9b correspond to those in Figure 9 of \cite{Stefani2021},
except that the abscissa is changed from period to 
frequency to ease the comparison with the sediment data.
When using only the positions of Jupiter and Saturn for computing the
Sun's orbital angular momentum (green curve), we basically obtain 
the dominant $22.14 \times 19.86/(22.14-19.86)=192.85\approx 193$\,years
period and its second and third harmonics at $\approx 96$ and $\approx 64$\,years,
plus one minor peak at $89.44 \approx 89$\,years which corresponds to the beat
period of Saturn's orbital period (29.424\,years) with the Hale period 
of 22.14\,years. 
However, when taking into account all planets (blue curve) 
we obtain, first of all,
three additional peaks at 82.7 years (beat period between 
the Jupiter-Neptune synodic period of 12.78\,years with 11.07\,years),
at 57.8\,years (beat period of the 
 Saturn-Neptune synodic period of 35.87\,years with 22.14\,years),
 and at 55.8\,years (beat period of the 
 Jupiter-Uranus synodic period of 13.91\,years with 11.07\,years).
 In addition to that, we observe the emergence of
 a third subharmonic of 193\,years at 579\,years, and 
 three fourths of 193\,years at 144\,years, both of which were 
 absent for  the pure Jupiter-Saturn forcing  
 (at present, we do not have a simple explanation for this
 somewhat counter-intuitive behaviour). To this, we add a third 
 (red) curve 
 which is distinguished from the blue one only by the 
 presence of some weak noise (with noise intensity $D=0.01$ 
 according to the notation \cite{Stefani2021}).
 Adjacent to this red curve, we summarize all the periods as 
 they were just discussed in connection with the green and blue 
 curves.
 
 When comparing the numerical results for the solar dynamo (Figure 9b) 
 with the climate-related observations (Figure 9a), the most
 striking feature is the nearly perfect agreement of the
 dominant and rather sharp Suess-de Vries period.
 The - still reasonable - agreement for the Gleissberg-type
 periods around 90\,years and slightly below 60\,years 
 becomes more complicated by the existence of some close-by peaks 
 (indicated by the dashed red lines) in the numerical model. Interestingly 
 enough, a decent correspondence  between the periodicities around 
 144 and 579 years can also be seen.
 Although we do not claim perfect accordance between our model and
 the observations, we find the match between the peaks, 
 in particular the sharp Suess-de Vries period, quite 
 remarkable.

\begin{figure}
\includegraphics[width=0.99\textwidth]{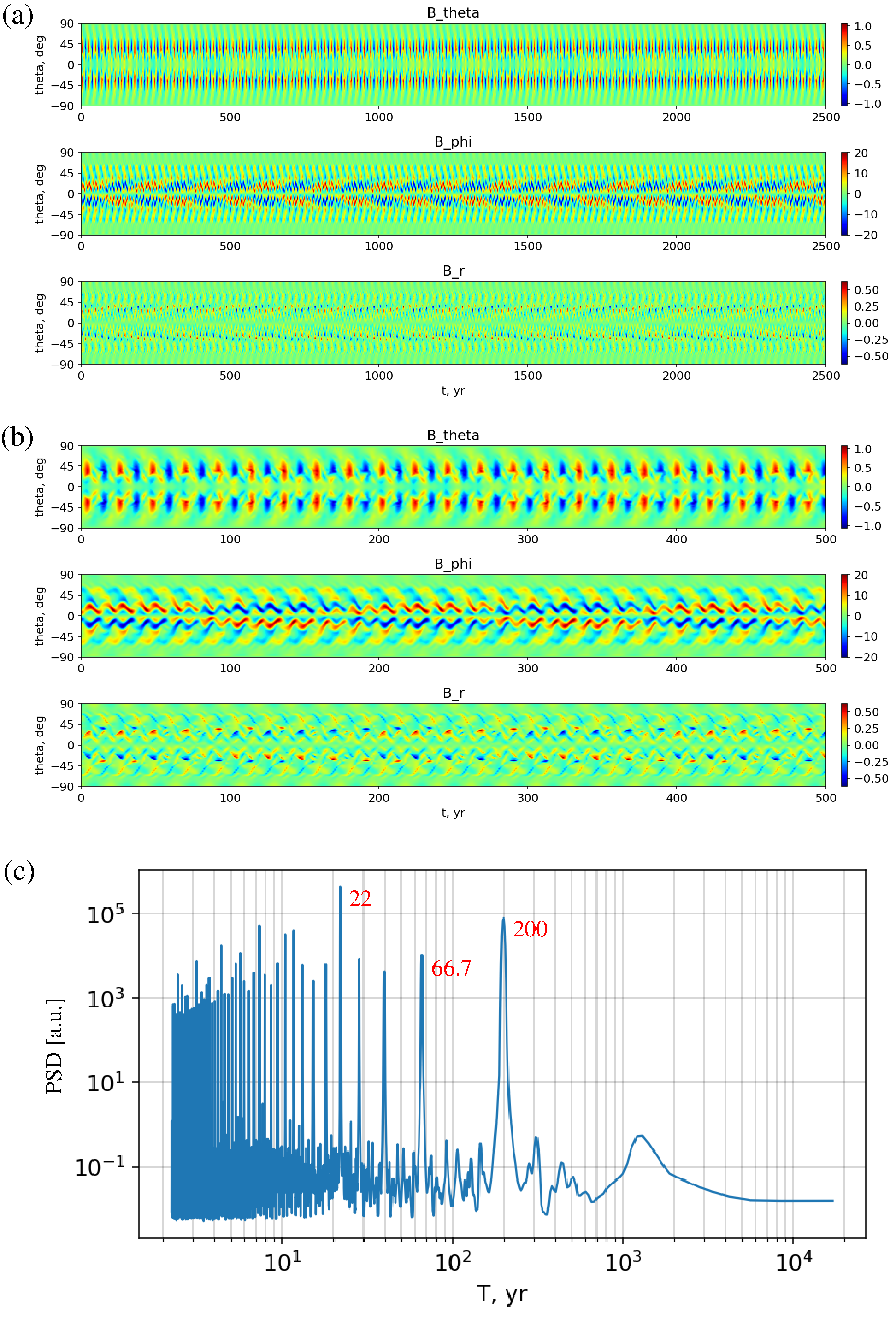}
  \caption{Appearance of a beat period of 200 years from a basic
  Hale cycle of 22 years and an 19.82-year period related to the
  barycentric 
  motion of the Sun. (a) Three magnetic field components at $r=0.7$ during a
  2500-year interval. (b) Details of (a) for a 500-year interval. (c) 
  PSD over the period $T$, with a clear dominance of the 22-year cycle,
  a subdominant 200-year cycle, and a Gleissberg-type cycle with 66.7 years.
  Note that the actual solar values are 22.14-years, 19.86 years, and 193-years.}
  \label{Fig:fig10}
\end{figure}

In the following, we go beyond 
this simple 1D model 
and try to confirm the emergence of the
Suess-de Vries cycle in frame of the 2D $\alpha-\Omega$-model 
as utilized by  \cite{Klevs2023}. In this model the {\it primary}
entrainment
of the Schwabe cycle was accomplished by a periodic, and
tidally synchronized, contribution
to the $\alpha$-effect in the tachocline region,
having the form of 
\begin{eqnarray} 
  \alpha^p(r,\Theta,t)&=&C^p_{\alpha} \frac{1}{\sqrt{2}}\sin^2 \Theta \cos \Theta 
  \left[1+{\rm erf} \left( \frac{r-r_c}{d} \right)  \right]
  \left[1-{\rm erf} \left( \frac{r-r_d}{d} \right)  \right] \times \nonumber \\  
  &&  \times \frac{2 |{\bf{B}}(r,\Theta,t)|^2}{1+|{\bf{B}}(r,\Theta,t)|^4}  \sin(2 \pi t/T_f  )\;,
   \end{eqnarray}
with $r_c=0.75$, $r_d=0.75$, $d=0.02$ (r is the dimensionless radius). Here the
resonance term has its maximum value equal to one 
at $|{\bf{B}}|=1$
which we re-interpret now as an
optimum field strength 
for the excitation of the three magneto-Rossby waves
whose beat period forms the 11.07-year envelope seen in Figure 8a.

In the 1D dynamo models of previous works \citep{Stefani2020a,Stefani2021}, 
the secondary 
19.86-year period of the barycentric  motion was implemented as 
an additional $B^3$-term in the induction equation representing a
periodically changing field-storage capacity of the 
tachocline. For the sake of simplicity, and in 
order to maintain  
the code structure of \cite{Klevs2023}, 
we decided to 
implement this effect as a time-periodic change 
of the optimal excitation condition
of the $\alpha$-effect in Equation (7), with 
the underlying (if vague) assumption
that any change of the differential rotation
(as discussed, e.g., by  
\cite{Zaqa1997,Javaraiah2003,Shirley2006,Sharp2013})
would also modify the resonance condition of the waves
\citep{Gachechiladze2019}).
Specifically, we replace the simple resonance 
term in the second line of Equation (7) by
the more complicated one
\begin{eqnarray} 
   \frac{2 (1+\epsilon
  \cos(2 \pi t /T_B))^2|{\bf{B}}(r,\Theta,t)|^2}{1+(1+\epsilon
  \cos(2 \pi t /T_B))^4|{\bf{B}}(r,\Theta,t)|^4}  \sin(2 \pi t/T_f  )
  \nonumber
   \end{eqnarray}
which keeps the amplitude of the prefactor at 1 but oscillates the 
position of this maximum between $1/(1+\epsilon)$ and $1/(1-\epsilon)$.

The results of the simulation for  the particular case 
$\epsilon=0.9583$
are shown in Figure 10. We stick here 
to the simplified Schwabe period $T_f=11.00$\,years, 
as in \cite{Klevs2023}, 
and use accordingly a
slightly modified value $T_B=19.82$\,years
of the barycentric period, in order to obtain 
also a simplified
beat period of $11.00 \times 19.82/(19.82-11)=200$\,years.

We see 
that this beat period of 200 years indeed shows up in all field
components, measured here at $r=0.7$.
We can identify, in particular, a clear 
Gnevyshev-Ohl rule \citep{Gnevyshev1948}
in terms if a succession of weak and strong cycles, which 
is changing its order every 200 years, as also observed 
for the solar cycle
by \cite{Tlatov2013}\footnote{Similar changes, on the Suess-de Vries time-scale, of 
the North-South asymmetry were recently reported by \cite{Mursula2023}.}.

We should note, however, that for the particular model
as chosen here such a clear behaviour occurs only in a not too
broad range of $\epsilon$, and that more complicated behaviours
are observed for other values.  A more detailed investigation of the
corresponding parameter dependencies, as well as the implementation of 
all planets in the angular momentum term, must be left for future studies.

\section{Bray-Hallstatt?}

Usually, solar activity variations with time scales of 1-3 kyr are
discussed under the name of Eddy and Bray-Hallstatt cycles
\citep{Steinhilber2012,Abreu2012,Soon2014,Scafetta2016,Usoskin2016}. 
However, it is by 
far not obvious whether the
very concept of ``cycles'' can  be 
transferred at all from the decadal 
(Schwabe, Hale) and centennial (Gleissberg, Suess-de Vries) 
to the millennial time
scale\footnote{Note that the underlying $^{14}$C and $^{10}$Be 
data bases have typical
durations of only 10 kyr, or just slightly longer,
see 
\cite{Kudryavtsev2020}}. In this context, 
the analyses of certain ice drift proxies
by \cite{Bond1997,Bond1999} 
had provided  evidence for
millennial climate
variability to occur in certain 1-3 kyr ``cycles'' of abrupt changes 
of the North Atlantic's surface hydrography which are 
closely related to corresponding grand minima of the
solar dynamo  \citep{Bond2001}.

\begin{figure}
\includegraphics[width=0.9\textwidth]{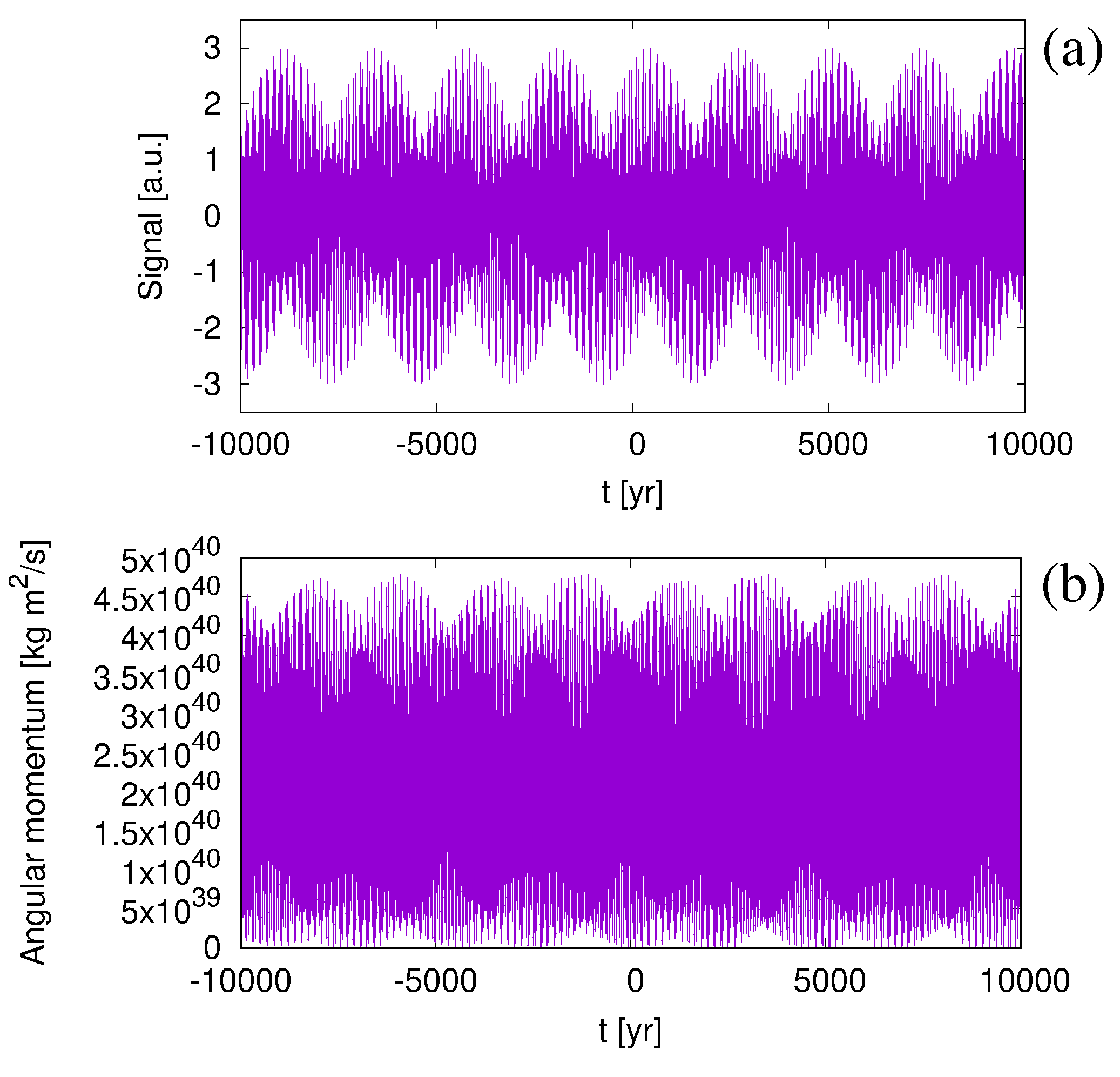}
  \caption{(a) Signal according to Equation (7) in the interval
  -10000-10000. (b) Orbital angular momentum of the Sun
  around the solar system's barycenter in the same interval,
  based on the DE431 ephemerides \cite{Folkner2014}.
  Note in either case the appearance of a 2318-yr period.}
  \label{Fig:fig11}
\end{figure}

\cite{Stefani2021} had shown a general tendency of the 
 Suess-de Vries cycle to undergo, for stronger forcings 
 of the 19.86-year term, abrupt breakdowns 
 which resemble these grand-minima, or Bond events.
Yet, such chaotic events do not exclude some remaining
regularity, perhaps driven by the 
2318-year cycle of the Jupiter-Saturn-Uranus-Neptune system,
according to Equation (37) of \cite{Scafetta2022},
\begin{eqnarray}
    s(t)=  \sin\left( 2\pi  \cdot \frac{t-t_{\rm JS}}{ P_{\rm JS}} \right) +\sin\left( 2\pi  \cdot \frac{t-t_{\rm SU}}{P_{\rm SU}} \right)
    +\sin\left( 2\pi  \cdot \frac{t-t_{\rm SN}}{ P_{\rm SN}} \right)
\end{eqnarray}
with the two-planet synodic periods $P_{\rm JS}=19.8593$\,years, 
$P_{\rm SU}=45.3636$\,years,
$P_{\rm SN}=35.8697$\,years, and the 
epochs of the conjunctions
$t_{\rm JS}=2000.48$,
$t_{\rm SU}=1988.44$, and
$t_{\rm SN}=1989.54$.

This function is visualized in Figure 11a, which corresponds (apart from the 
enlarged time interval)
to Figure 5E in  \cite{Scafetta2022}.
Less formal than this equation, though, is the actual orbital 
angular momentum, shown in Figure 11b, which we have computed 
from the DE431 ephemerides described in \cite{Folkner2014}.
Some obvious phase-shift notwithstanding, the same 2318-yr periodicity 
shows up as in Figure 11a,
which indeed speaks for the dominant role of the 
two-planet synodic periods as indicated by Scafetta's equation (7).
It remains to be seen in future work if the implementation
of this real angular momentum curve into the extended 
2D $\alpha-\Omega$ dynamo model of Section 5 leads to any
noticeable signal on that time-scale, too. 

Actually, such a  2300\,year modulation had been found in 
proxies of the galactic cosmic radiation by \cite{Mccracken2008}.
Interestingly, during the observed 
cosmic radiation enhancements these authors also noticed 
a strengthened Suess-de Vries cycle (their 
Figure 3), which is somehow consistent with a similar numerical
result in Figures 11 and 12 of \cite{Stefani2021}. This
entire behavior is reminiscent of the concept of 
{\it supermodulation} as described by 
\cite{Weiss2016}.

As a side remark, note that a similar interplay of chaos and 
regularity had been discussed as a {\it stochastic resonance} phenomenon in 
connection with the reversal statistics of the geodynamo for which 
the 95-kyr Milankovic
cycle of Earth's orbit excentricity seems to play 
a decisive role \citep{Consolini2003,Stefani2006,Fischer2008}.

\section{Summary and conclusions}

In this paper, we have pursued the programme of explaining the various 
temporal variabilities of solar activity as being triggered by gravitational 
influences of the orbiting planets. In doing so, we
focused not only on what the Sun is being 
``told'' by the the planets, 
but also on what it is able to ``understand''.
On each timescale involved we have therefore asked about the relevant
intrinsic processes within the Sun that might be able 
to resonate with the respective external forcings.

Starting from the
shortest, Rieger-type timescales  of some 100-300 days, we 
have shown that the two-planet spring tides of the 
tidally dominant planets Venus, Earth and Jupiter are indeed capable 
of exciting magneto-Rossby waves with amplitudes of up to m/s, or 
even more, depending on the widely unknown damping parameter
$\lambda$.
This identification of a mechanism by which 
the tidal energy can
be ``harvested'' by the Sun reinstates the relevance of the 
1\,mm tidal height, which has been known for a long time to 
correspond energetically to a velocity of 
1\,m/s \citep{Opik1972}, but which was often 
(mis-)used to deride any sort of solar-dynamo 
synchronization.

We went on by asking how the beat periods of 
those magneto-Rossby waves that are excited by the two-planets'
spring tides might be capable of synchronizing a conventional
$\alpha-\Omega$ dynamo. While it had been shown previously 
\citep{Klevs2023} that an $\alpha$-effect 
in the range
of dm/s could lead to parametric resonance
of an underlying conventional $\alpha-\Omega$ dynamo, 
we focused here on the two most promising mechanisms, viz, the 
sensitivity of field storage capacity and helicity oscillations. 
In either case we argued that it is the 
nonlinear (basically squared) action 
of the tidally triggered waves that is important, 
which naturally introduces the  11.07-year beat period 
period.
We have to admit, however, that the exact mechanism is not 
yet understood and will require more investigations in the 
future. 
The most obvious next step in this direction is the
computation of the $\alpha$-effect, utilizing the
procedures of \cite{Avalos2009}.

A similar caveat applies to the second mechanism which, again, 
tries to explain long-term periods by the beats 
of shorter-term periods, 
this time focusing on the Suess-de Vries cycle.
We have used a slightly enhanced version of the
2D-dynamo model of \cite{Klevs2023} 
to explain the latter as a beat period 
between the 22.14-year Hale cycle and the 
19.86-year period of the 
rosette-shaped barycentric motion of the Sun.

In our view, this conceptual shift towards first 
identifying the energy transfer on the shortest-possible time 
scales and looking afterwards for beat periods of the excited 
waves solves a couple of problems, among them
the lack of direct tidal forcing at the 11.07-yr period
\citep{Nataf2022,Cionco2023} as well as Charbonneau's  
noise-related argument against the viability of a 
{\it direct} planetary forcing on the centennial and millennial
time-scale \citep{Charbonneau2022} .

In this respect, our line of argument is quite similar to that of 
\cite{Raphaldini2019,Raphaldini2022}
who also tried to explain, first, the Schwabe cycle in terms of 
nonlinear triads of magneto-Rossby waves with a ten-times 
shorter time-scale, and, second, the Gleissberg and Suess-de Vries 
cycles as higher order ``precession resonances'', 
including two such triads.
In contrast to these works, however, we focused
here on the very specific 11.07-year period that shows up in the 
non-linear interaction of three 
tidally-induced magneto-Rossby waves. 
A second difference applies to the transition from Rieger, via Schwabe,
to Suess-de Vries for which, in our opinion, the argument of
\cite{Raphaldini2022} may run into conceptual problems due 
to the intervening oscillatory dynamo field which 
changes ``on-the-fly'' 
the eigenfrequencies of the underlying magneto-Rossby waves. 

By contrast, we relied here on an {\it independent beat mechanism} 
between the 
22.14-year Hale cycle and the 19.86-year barycentric motion which 
has the absolutely strongest effect on the angular momentum (see
Figure 2 in \cite{Stefani2021}), although the 
specific spin-orbit coupling mechanism is not well 
understood yet. Again, our 
focus was on a sharp 193-year beat period (as indeed 
observed by \cite{Prasad2004}) rather than on 
some broad signal on the general 
100--200-year scale.
That said, we believe that the mathematical framework of 
triadic (and ``precession'') interactions as developed by
\cite{Raphaldini2019,Raphaldini2022}, together with the
methods to derive the $\alpha$-effect for Rossby waves
\cite{Avalos2009}, represent
an ideal starting point for more quantitative corroborations 
of the ideas that could only be sketched in this article.
Complementarily to this focus on the $\alpha$-effect,
one might also evaluate  the  non-linear energy transfer 
terms from the Appendix of 
\cite{Dikpati2018} which, via tachocline nonlinear oscillations,
may lead to a similar $\Omega$-related 
synchronization mechanism as
once proposed by \cite{Zaqa1997}.

As for the longest time scales (Eddy, Bray-Hallstatt) we 
have reiterated the argument of \cite{Stefani2021} that they 
might well occur as chaotic
breakdowns of the shorter Suess-de Vries cycle, quite in accordance with 
the supermodulation concept
of \cite{Weiss2016}. Yet, this does not exclude
some sort of residual regularity, perhaps driven by the 
2318-year cycle of the Jupiter-Saturn-Uranus-Neptune system
for which there is also some observational 
evidence  \citep{Mccracken2008}.

In view of the growing evidence for a 
significant role of the Sun for the terrestrial climate,
both on the decadal and centennial \citep{Connolly2021,Stefani2021b,Scafetta2023} 
as well as on the millennial time-scale \citep{Bond2001},
we consider a deepened understanding of any such kind of
(quasi-)deterministic triggers of the solar dynamo as 
worthwhile and timely.

\section*{Appendix}

In this Appendix, we are concerned with two specific 
aspects related to the azimuthal averages carried out in Section 4.

First, we explain why the azimuthally integrated 
square of the 3-wave-signal still contains the 11.07-year periodicity
(Figure 8a),
while that of the corresponding 2-wave-signal (Figure 8b) does not.
For that, we show in Figure 12 the corresponding 
squares of the signals 
at different angles $\varphi$ (indicated by the colored
numbers).
Obviously, for the 3-wave signal (a) the maximum of the beat for 
different angles remains at the same instant in time 
(reflecting the ``orbital invariance'' according to \cite{Scafetta2022}),
while it shifts linearly for the 2-wave signal (b).
The integration over the angles explains the 
differences seen
in Figure 8.

\begin{figure}
\includegraphics[width=0.80\textwidth]{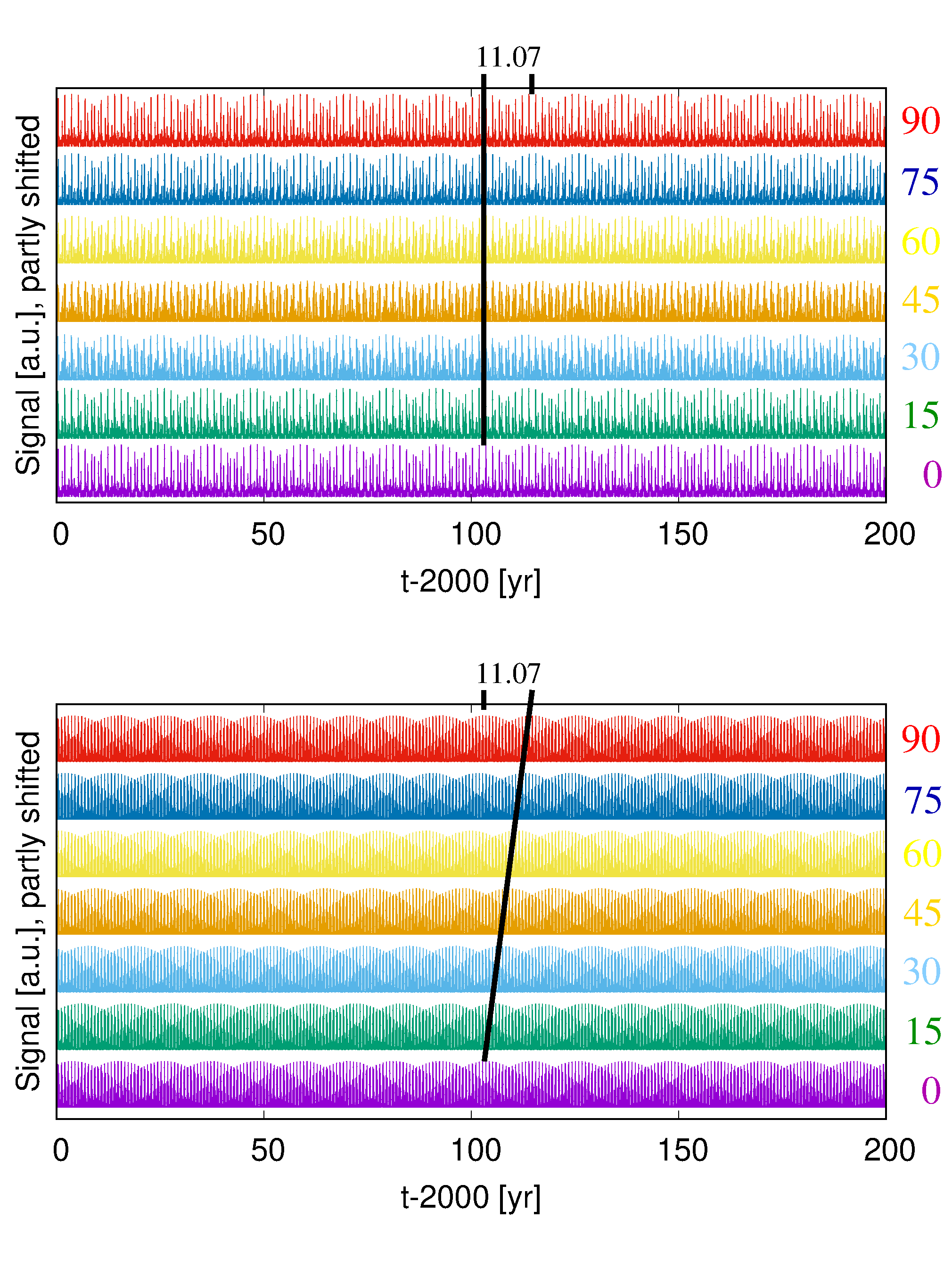}
  \caption{Values of the squares under the integrals
  of Equation (6) at 7 different angles $\varphi$ (colored number on the r.h.s.), 
  for the entire 3-wave signal (a) and the 2-wave signal 
  without the last term (from Venus-Earth).
  The black line links the instants of the 
  maxima at different angles. While this instant remains
  the same in (a), it is shifted to later instants in (b).}
  \label{Fig:fig12}
\end{figure}

Second, we consider the issue that the dynamo
relevant non-linear term does not correspond exactly 
to Equation (6)
but to  a more complicated one including an { \it additional} 
phase shift $\psi_0$ in the second factor, according to

\begin{eqnarray}
    S_{\rm ps}(t)&=& \frac{1}{2 \pi} \int_0^{2 \pi} d\varphi
    \left[ \cos\left( 2\pi  \cdot \frac{t-t_{\rm VJ}}{0.5 \cdot P_{\rm VJ}} +2 \varphi \right) +\cos\left( 2\pi  \cdot \frac{t-t_{\rm EJ}}{0.5 \cdot P_{\rm EJ}} +2 \varphi\right) \right.  \nonumber \\
    &&\,\,\,\,\,\,\,\,\,\,\,\,\,\,\,\,\,\,\,\,\,\,\,\,\,\,\,\,\,\,\,\,\,\,\,\,
    +\left.  \cos\left( 2\pi  \cdot \frac{t-t_{\rm VE}}{0.5 \cdot P_{\rm VE}} +2 \varphi  \right) \right] \nonumber \\
    &&\times
    \left[ \cos\left( 2\pi  \cdot \frac{t-t_{\rm VJ}}{0.5 \cdot P_{\rm VJ}} +2 \varphi +\psi_0 \right) +\cos\left( 2\pi  \cdot \frac{t-t_{\rm EJ}}{0.5 \cdot P_{\rm EJ}} +2 \varphi +\psi_0 \right) \right.  \nonumber \\
    &&\,\,\,\,\,\,\,\,\,\,\,\,\,\,\,\,\,\,\,\,\,\,\,\,\,\,\,\,\,\,\,\,\,\,\,\,
    +\left.  \cos\left( 2\pi  \cdot \frac{t-t_{\rm VE}}{0.5 \cdot P_{\rm VE}} +2 \varphi +\psi_0 \right) \right]
\end{eqnarray}

Such a phase shift would be relevant, in particular, for 
the azimuthally averaged 
helicity wherein - for the case of
magneto-Rossby waves - the vorticity  is typically phase 
 shifted by 90$^{\circ}$ with respect to the vertical velocity.

Although it is rather trivial in view of the
integration law of products of two trigonometric functions, 
Figure 13 shows how this average converges to zero with increasing
phase shift $\psi_0$ (illustrated by $\psi_0$ 
values of 0$^{\circ}$, 45$^{\circ}$ and
75$^{\circ}$.)

\begin{figure}
\includegraphics[width=0.80\textwidth]{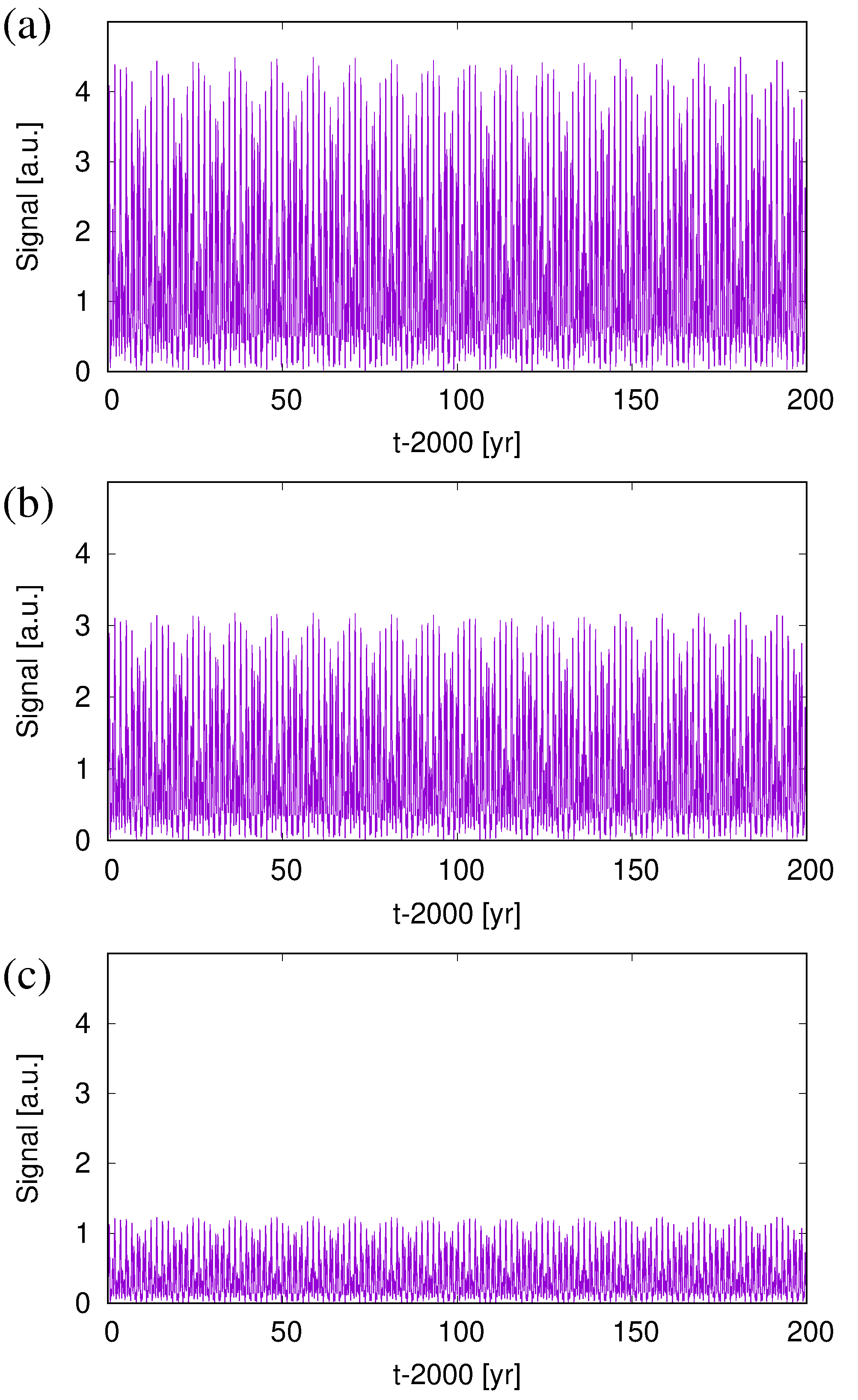}
  \caption{Signals $S_{\rm ps}$ 
  according to Equation (9) with the phase shifts
  $\psi_0$ (a) equal to 0$^{\circ}$, (b) equal to 45$^{\circ}$, (c) equal to 
  75$^{\circ}$.}
  \label{Fig:fig13}
\end{figure}

\begin{acks}
This work received funding from the European Research Council 
(ERC) under the European Union's Horizon 2020 research and innovation 
programme (grant agreement No 787544).  Inspiring discussions with 
Carlo Albert, J\"urg Beer, Axel Brandenburg, Robert Cameron, 
Antonio Ferriz Mas, G\"unther R\"udiger, Laur\`ene Jouve, Henri-Claude Nataf, Markus Roth, 
Dmitry Sokoloff, Rodion Stepanov, 
Steve Tobias, and Teimuraz Zaqarashvili on various aspects of the 
solar dynamo, and 
its synchronization, are gratefully acknowledged.
\end{acks}

\section*{Disclosure of Potential Conflicts of Interest}
The authors declare that they have no conflicts of interest.

\end{article} 

\end{document}